\documentclass[letterpaper]{article} 
\usepackage{aaai2026} 
\usepackage{times} 
\usepackage{helvet} 
\usepackage{courier} 
\usepackage[hyphens]{url} 
\usepackage{graphicx} 
\usepackage{natbib} 
\usepackage{caption} 
\usepackage{algorithm}
\usepackage{algorithmic}
\usepackage{newfloat}
\usepackage{listings}
\DeclareCaptionStyle{ruled}{labelfont=normalfont,labelsep=colon,strut=off} 
\floatstyle{ruled}
\newfloat{listing}{tb}{lst}{}
\floatname{listing}{Listing}
\usepackage{booktabs}
\usepackage{xcolor}
\usepackage{enumitem}
\usepackage{amsmath}
\usepackage{amssymb}
\usepackage{subfig}
\usepackage{multirow}
\usepackage[most]{tcolorbox}
\usepackage{fontawesome}
\usepackage[hidelinks]{hyperref}

\newtcolorbox{prompt}{
 colback=gray!5, 
 colframe=gray!25, 
 arc=2mm, 
 boxsep=0pt, 
 left skip=0pt, 
 right skip=0pt, 
 width=\linewidth, 
}

\newcommand{\methodname}{\textsc{ERank}}
\title{\methodname: Fusing Supervised Fine-Tuning and Reinforcement Learning for Effective and Efficient Text Reranking}
\author{
 Yuzheng Cai\textsuperscript{\rm 1},
 Yanzhao Zhang\textsuperscript{\rm 2}, 
 Dingkun Long\textsuperscript{\rm 2}, 
 Mingxin Li\textsuperscript{\rm 2}, 
 Pengjun Xie\textsuperscript{\rm 2}, 
 Weiguo Zheng\textsuperscript{\rm 1}\thanks{Corresponding author}
}
\affiliations{
 \textsuperscript{\rm 1}Fudan University \textsuperscript{\rm 2}Alibaba Group\\
 yuzhengcai21@m.fudan.edu.cn,
 zhengweiguo@fudan.edu.cn
\\
{\tiny $\quad$}
\\
\href{https://huggingface.co/collections/Alibaba-NLP/erank-68b302be7d1f62b0b7cb19a2}{\includegraphics[height=3.1mm]{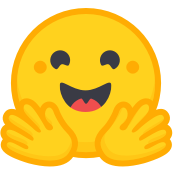} HuggingFace}
\quad
\href{https://modelscope.cn/collections/ERank-488bc43c873e4c}{\includegraphics[height=3.3mm]{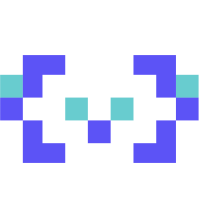} ModelScope}
}

\begin{document}

\maketitle

\begin{abstract}
Text reranking models are a crucial component in modern systems like Retrieval-Augmented Generation, tasked with selecting the most relevant documents prior to generation. However, current Large Language Models (LLMs) powered rerankers often face a fundamental trade-off. On one hand, Supervised Fine-Tuning based pointwise methods that frame relevance as a binary classification task lack the necessary scoring discrimination, particularly for those built on reasoning LLMs. On the other hand, approaches designed for complex reasoning often employ powerful yet inefficient listwise formulations, rendering them impractical for low latency applications. To resolve this dilemma, we introduce {\methodname}, a highly \textit{Effective} and \textit{Efficient} pointwise reranker built from a reasoning LLM that excels across diverse relevance scenarios. We propose a novel two-stage training pipeline that begins with Supervised Fine-Tuning (SFT). In this stage, we move beyond binary labels and train the model generatively to output fine grained integer scores, which significantly enhances relevance discrimination. The model is then further refined using Reinforcement Learning (RL) with a novel, listwise derived reward. This technique instills global ranking awareness into the efficient pointwise architecture. We evaluate the {\methodname} reranker on the BRIGHT, FollowIR, TREC DL, and BEIR benchmarks, demonstrating superior effectiveness and robustness compared to existing approaches. On the reasoning-intensive BRIGHT benchmark, our {\methodname}-4B achieves an nDCG@10 of $38.7$, while a larger 32B variant reaches a state of the art nDCG@10 of $40.2$.
\end{abstract}

\section{Introduction}
\label{sec:intro}

\begin{figure}[t]
 \centering
 \includegraphics[width=\linewidth]{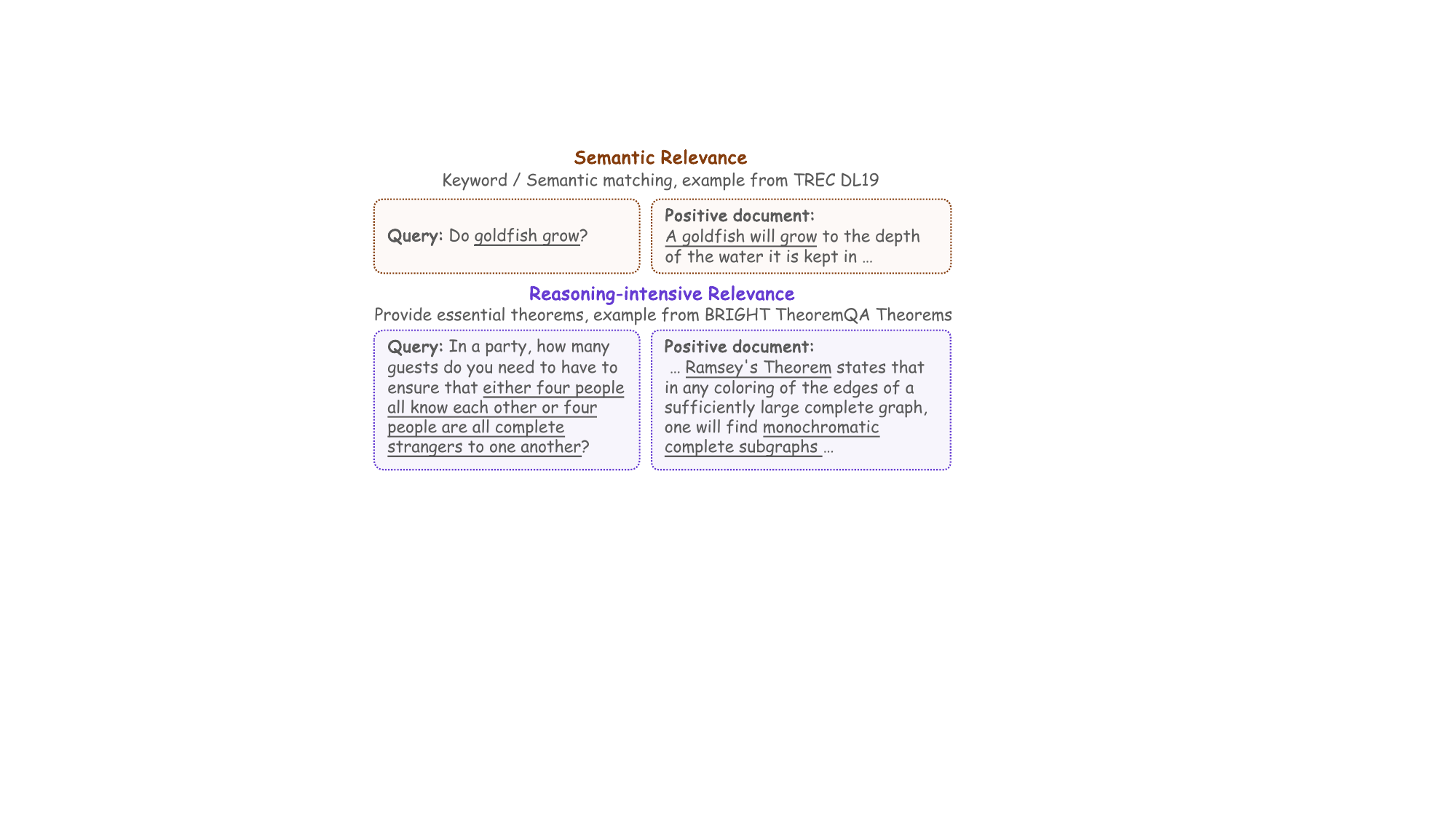}
 \caption{Semantic relevance refers to the traditional understanding based on keyword or semantic matching, while the reasoning-intensive example aims to capture documents that may not directly answer the query but provide essential intermediate information needed for multi-step reasoning.}
 \label{fig:relevance_small}
\end{figure}

\begin{figure}[!htp]
 \centering
 \includegraphics[width=\linewidth]{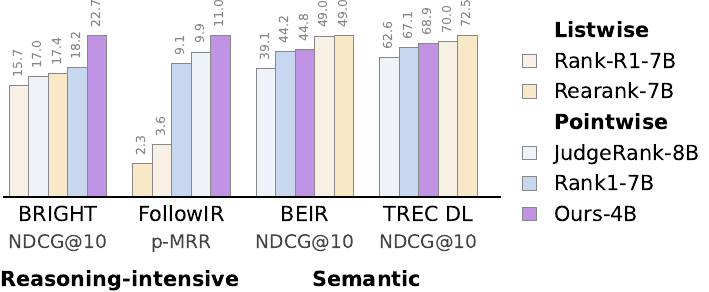}
 \caption{{\methodname}-4B achieves state-of-the-art performance among pointwise rerankers using candidate documents retrieved by BM25 with original queries.
 Under retrieval settings in Section~\ref{sec:implementation}, {\methodname}-4B and 32B further achieve the nDCG@10 of 38.7 and 40.2 on BRIGHT, respectively.}
 \label{fig:effectiveness}
\end{figure}

Text reranking is a fundamental component of various Natural Language Processing and Information Retrieval applications, utilized extensively in downstream tasks such as open-domain question answering~\cite{lee2018ranking}, web search~\cite{lin2022pretrained}, and recommendation systems~\cite{chuang2020tpr,gao2025llm4rerank}. Large Language Models (LLMs) have significantly reshaped the text reranking landscape. On one hand, studies have sought to leverage the advanced text understanding capabilities of LLMs for reranking, either through zero-shot prompting or Supervised Fine-Tuning~\cite{zhang2023empirical,liu2024information}. On the other hand, LLMs have introduced new application paradigms like Retrieval-Augmented Generation~\cite{RAG-survey-1,RAG-survey-2,RAG-survey-3} and agentic systems~\cite{huang2024understanding,li2024survey}. These paradigms demand capabilities beyond traditional semantic relevance, requiring models to perform reasoning-intensive retrieval, such as identifying issue-relevant code snippets to resolve a specific programming problem.
Recent advancements in test-time compute~\cite{OpenAI-O1, DeepSeek-R1} have shown promise in such scenarios, with a growing number of text rerankers based on reasoning LLMs~\cite{Rank1, Rank-R1, Rearank}.

Prior approaches have generally treated traditional semantic relevance and reasoning-intensive reranking as distinct challenges, which are illustrated in Figure~\ref{fig:relevance_small}. For semantic tasks, Supervised Fine-Tuning (SFT) is a common strategy~\cite{repLLaMA,RankGPT,zhang2024two}. However, most SFT-based rerankers adopt a pointwise scoring method based on binary classification, where the model predicts labels like ``Relevant'' or ``Not Relevant''. We argue this approach is suboptimal as it leads to poor score discrimination, a problem exacerbated in modern reasoning LLMs that generate overconfident predictions after Chain-of-Thought (CoT). For reasoning-intensive tasks, Reinforcement Learning (RL) has shown promise~\cite{Rank-R1,Rearank}. However, these methods often rely on listwise or setwise formulations and ingest multiple candidate documents simultaneously.
With sliding windows, they process different batches of documents sequentially, resulting in prohibitive latency and memory footprints that make them impractical for real-world deployment.

\begin{figure*}[t]
 \centering
 \includegraphics[width=\linewidth]{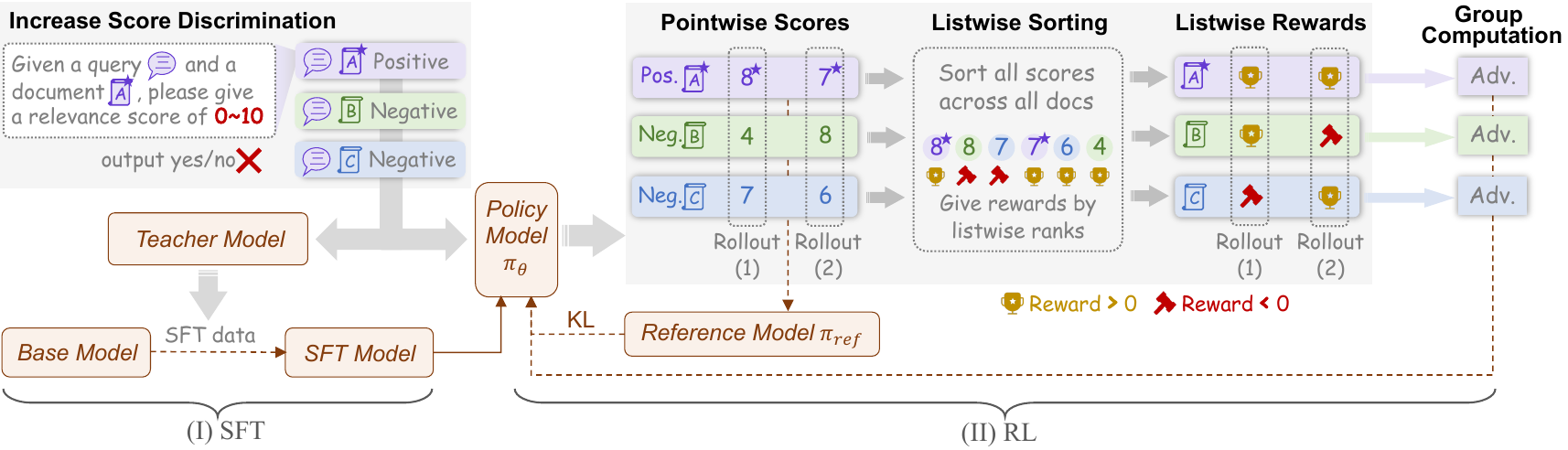}
 \caption{Overview of the two-stage fine-tuning pipeline for the pointwise {\methodname} reranker.
 Given a query and $N=3$ documents A, B and C, where A is the only positively related one, the SFT model is trained to deliver a relevance score ranging from $0$ to $10$ for each document.
 During RL training, the model generates $G=2$ rollouts for each document.
 These $N \times G=6$ scores extracted from all rollouts across all documents are then sorted together to compute listwise ranking derived rewards.
 }
 \label{fig:overview}
\end{figure*}

This work addresses a central question: can a single, efficient reranker powered by reasoning LLM be trained to excel at both semantic relevance and deep reasoning? We contend that this is achievable by enhancing the pointwise architecture, which scores each document independently. We introduce a novel, two-stage training framework illustrated in Figure~\ref{fig:overview}, which seamlessly integrates Supervised Fine-Tuning (SFT) with Reinforcement Learning (RL) for LLM-based reranker training. The first stage, SFT, trains a base model on a diverse mixture of semantic and reasoning-oriented data. Crucially, we abandon the standard binary classification paradigm and instead train the model using a fine-grained integer scoring scheme, which fully utilizes the generative power of LLMs and significantly improves score discrimination. We also employ a data synthesis strategy to generate high-quality reasoning chains and fine-grained scores to overcome data scarcity. In the second stage, we further refine the SFT-tuned model using RL. To bridge the gap between listwise optimality and pointwise efficiency, we introduce a novel, listwise-derived reward function. This function provides a global ranking signal during training, encouraging the model to learn the relative importance of documents. This allows our pointwise model to benefit from listwise-style optimization while retaining its low latency. 

\begin{figure}[t]
 \centering
 \includegraphics[width=\linewidth]{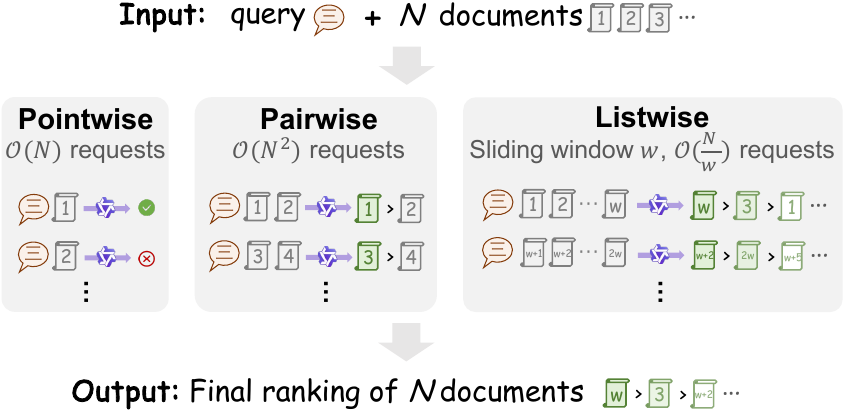}
 \caption{Comparison of different reranking paradigms.}
 \label{fig:paradigm}
\end{figure}

Extensive experiments on semantic (TREC DL~\cite{DL19, DL20}, BEIR~\cite{BEIR}) and reasoning-intensive benchmarks (BRIGHT~\cite{BRIGHT}, FollowIR~\cite{FollowIR}) confirm that our framework delivers substantial gains. As shown in Figure~\ref{fig:effectiveness}, our 4B-parameter model outperforms many 7B model size rerankers, and our 32B model sets a new state-of-the-art on the BRIGHT benchmark. Latency measurements confirm our models maintain the high efficiency of standard pointwise rerankers, making them both powerful and practical.

Briefly, our main contributions are to:
\begin{itemize}[leftmargin=*]
 \item Reveal the suboptimality of binary classification for LLM rerankers and propose a generative approach that outputs discrete integer scores to enhance score discrimination.
 \item Introduce a novel two-stage framework integrating Supervised Fine-Tuning (SFT) and Reinforcement Learning (RL) to build a single, efficient pointwise reranker for both semantic and reasoning-intensive tasks.
 \item Our model, {\methodname}, sets a new state-of-the-art on the reasoning-based BRIGHT benchmark while demonstrating exceptional performance on standard semantic tasks.
 \item We will open-source our trained models and data to facilitate reproducibility and future research.
\end{itemize}

\section{Related Work}
\label{sec:related_work}
\paragraph{LLM for Text Reranking.}
Large Language Models (LLMs) have significantly advanced text reranking beyond the capabilities of earlier encoder-only models such as BERT~\cite{liu2024information}. LLMs are typically applied to this task using either zero-shot prompting~\cite{zhang2023empirical,Setwise,JudgeRank} or, more effectively, Supervised Fine-Tuning~\cite{repLLaMA,zhang2024two}. As shown in Figure~\ref{fig:paradigm}, reranking methodologies are broadly categorized into pointwise~\cite{liang2022holistic}, pairwise~\cite{qin2024large}, and listwise approaches~\cite{ma2023zero,RankGPT}. Listwise methods, which evaluate a list of candidate documents, generally yield the highest ranking quality by directly optimizing the document order~\cite{rethinkBERT,zhang2022hlatr,liu2025listconranker}. However, their computational cost scales quadratically with input length, making them impractical for real-world systems that demand low latency. In contrast, pointwise methods score each query-document pair independently. This paradigm enables massive parallelization and efficient inference, establishing it as the preferred choice for large scale deployment. Most fine-tuned pointwise rerankers conventionally treat the task as a binary classification problem. We argue this approach fails to leverage the full generative power of modern LLMs and results in suboptimal performance.

\paragraph{Reinforcement Learning for Reranking.}
The success of Reinforcement Learning (RL) in enhancing the complex reasoning abilities of LLMs, exemplified by models like OpenAI-O1~\cite{OpenAI-O1} and DeepSeek-R1~\cite{DeepSeek-R1}, has inspired its application to reranking. Recent work demonstrates that RL can refine a model's capacity to identify documents that are not merely semantically relevant but also instrumentally useful for resolving a user's query. However, these pioneering RL-based reranking methods predominantly adopt listwise or setwise training frameworks~\cite{Rank-R1,Rearank}. While effective, they inherit the high latency and memory requirements associated with processing multiple batches of documents sequentially, which limits their practical applicability. Our work addresses this critical gap. We introduce a novel two-stage training pipeline that begins with a generative Supervised Fine-Tuning stage using fine-grained scoring to improve relevance discrimination. Subsequently, we employ RL to optimize our efficient pointwise model with a globally aware listwise reward signal. This strategy achieves the ranking quality of listwise methods while preserving the inference efficiency of a pointwise architecture.

\section{Method}
\label{sec:method}
Our training methodology unfolds in a two-stage pipeline designed to build a reranker that excels at both semantic relevance and reasoning-intensive relevance. The first stage uses Supervised Fine-Tuning (SFT) to establish a strong foundation, and the second stage employs Reinforcement Learning (RL) with the Group Relative Policy Optimization (GRPO) algorithm~\cite{GRPO} to refine the reranking ability.

\subsection{Task Formulation} 
We formulate the text reranking task as a generative process. With a specific instruction $I$ that defines the relevance criteria, given a query $q$ and a set of $N$ candidate documents $\{d_1, d_2, \dots, d_N\}$, our model processes each query-document pair independently. For each pair, it generates a response that includes a Chain-of-Thought (CoT) $c_i$ explaining its reasoning, followed by a final relevance score $s_i$. This is represented by the conditional probability of the policy LLM $\pi$:

{
\vspace{-3mm}
\begin{equation*}
\pi(c_i, s_i \mid I, q, d_i), \quad i = 1, 2, \dots, N.
\end{equation*}
}

Based on the extracted scores $\{s_1, s_2, \dots, s_N\}$, the documents are then sorted in descending order to produce the final ranked list. This pointwise formulation ensures low inference latency and provides interpretability through the generated reasoning Chain-of-Thought (CoT).

\subsection{Supervised Fine-Tuning with Fine-Grained Scores}
\label{sec:SFT}
To perform Supervised Fine-Tuning (SFT), we construct a dataset $\mathcal{D} = \{(x_k, y_k)\}_{k=1}^K$, where each input $x_k$ is a prompt containing the relevance instruction $I$, a query $q$, and a document $d$. The target output $y_k$ is a combination of reasoning chain $c$ and relevance score $s$, as formatted in Figure~\ref{fig:prompt_0_10}.

\paragraph{Generative Fine-Grained Scoring.}
A central limitation of prior pointwise rerankers is their reliance on a binary classification objective, where the model is trained to predict labels of \textit{``yes''} and \textit{``no''} to represent relevance. To better leverage the generative nature of Large Language Models (LLMs), some recent methods have exploited their autoregressive capabilities. These approaches compute a normalized relevance score by extracting token probabilities for \textit{``yes''} and \textit{``no''}, which can be definded as:

{
\vspace{-2mm}
\begin{equation*}
\frac{\text{Pr(token=\textit{yes})}}{\text{Pr(token=\textit{yes}) + Pr(token=\textit{no})}}.
\end{equation*} 
}

Our experiments reveal that such strategy leads to poor score discrimination. The issue becomes particularly pronounced when using reasoning LLMs, since the confidence-boosting effect of Chain-of-Thought (CoT) reasoning causes these models to generate overconfident predictions. As illustrated in Figure~\ref{fig:probs} and detailed in Appendix~\ref{app:preliminary_exp}, a comparison between a non-reasoning model (Qwen3-32B~\cite{Qwen3}) and a reasoning-enhanced model (QwQ-32B~\cite{QwQ}) shows that the latter produces normalized scores heavily concentrated near 0 or 1. A significantly higher proportion of scores from the reasoning model falls within the extreme intervals of $[0, 0.00001]$ and $[0.99999, 1]$. This concentration severely diminishes the model's ability to distinguish between varying degrees of relevance, which is essential for effective reranking.

To overcome this limitation, we reframe reranking as a generative task with a fine-grained scoring system. Instead of predicting binary labels, we train the model to generate an integer score from $0$ to $10$ that reflects the degree of relevance with prompt in Figure~\ref{fig:prompt_0_10}. Then, the final ranking score is computed as $s_i \times \text{Pr(token=}s_i)$. This method fully utilizes the autoregressive capabilities of the LLM, creating a more expressive and discriminative scoring space critical for distinguishing between documents of varying quality. Table~\ref{tab:discrimination} shows that it consistently improves nDCG@10 across multiple benchmarks under experimental settings in Appendix~\ref{app:preliminary_exp}.

\begin{figure}[t]
\begin{prompt}
\small
Given a query and a document, please give a relevance score of 0 to 10.
The goal or relevance definition is: \texttt{\{instruction\}}
\vspace{2mm}

Here is the query:\\
\texttt{\{query\}}
\vspace{2mm}

Here is the document:\\
\texttt{\{document\}}
\vspace{2mm}

After thinking, directly choose a relevance score from [0, 1, 2, 3, 4, 5, 6, 7, 8, 9, 10].\\
- 0 represents completely not related.\\
- 10 means perfectly related.
\vspace{2mm}

Desired output format:\\
\texttt{<think>}put your thinking here\texttt{</think>}\texttt{<answer>} Only allows an integer here\texttt{</answer>}
\vspace{2mm}

Your output:
\end{prompt}
\vspace{-2mm}
\caption{Prompt for scoring with integers from 0 to 10.}
\label{fig:prompt_0_10}
\end{figure}

\paragraph{Data Synthesis for SFT.}
To train our model for this fine-grained scoring task, we synthesize a high-quality dataset that covers both semantic matching and complex reasoning scenarios. We employ a powerful open-source model, QwQ-32B~\cite{QwQ}, as a teacher to generate reasoning chains and integer scores. To ensure the quality and reliability of the synthetic labels, our data construction process emphasizes two key aspects. \textbf{Query-Document Diversity}: We source query-document pairs from a diverse mix of datasets, including MS MARCO~\cite{MS-MARCO} for semantic relevance, and ReasonIR~\cite{ReasonIR} and Promptriever~\cite{promptriever} for complex reasoning tasks. For semantic relevance, we randomly select $5,000$ queries from the MS MARCO dataset. For reasoning-intensive tasks, we sample $10,000$ queries from the hard query (HQ) set of ReasonIR and $5,000$ queries from the Promptriever training set. Both of these sources contain complex queries that require deep reasoning. For each query, we enrich the initial candidate pool, which includes the annotated positive documents and synthetic negative documents, by retrieving the top $1,000$ documents from the corpus using the ReasonIR-8B retriever. We then sample documents from different ranking ranges to create a balanced set of negatives: the top $10$ documents serve as hard negatives, positions $11–100$ as medium negatives, and positions $101–1,000$ as easy negatives. Further details are provided in Appendix~\ref{app:train_data}. Each query is ultimately associated with exactly $20$ documents to maintain a consistent input structure. \textbf{High-Quality and Stable Reasoning Trajectory Generation}: We use the QwQ-32B teacher model to generate a reasoning chain $c$ and a corresponding score $s$ for each query-document pair. To improve the reliability of these generated labels, we perform multiple independent generations for each instance and compute the average score, which serves as a consensus score. We then select the single generation whose score is closest to this consensus. Experiments on a random sample of $512$ queries show that this strategy significantly improves scoring quality. The average nDCG@10 increased from $63.5\%$ with a single generation to $65.6\%$ with $3$-sample consensus and further to $67.4\%$ with $10$-sample consensus. To balance performance and computational cost, we use three generations per instance in our final data synthesis process. Finally, we filter out any instances where the generated output exceeds $2,048$ tokens. This procedure results in our final Supervised Fine-Tuning (SFT) dataset, $\mathcal{D}$.

\begin{figure}[t]
\centering
\includegraphics[width=\linewidth]{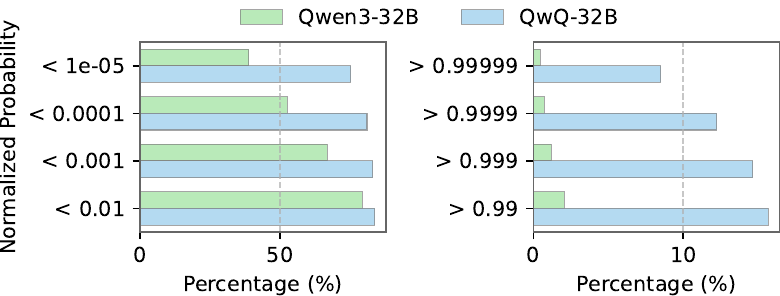}
\caption{Distributions of normalized probability with non-reasoning and reasoning LLMs on BRIGHT benchmark.}
\label{fig:probs}
\end{figure}

\begin{table}[t]
 \centering
 \small
 \begin{tabular}{c|ccc}
 \toprule
 \textbf{Scoring} & \textbf{BRIGHT} & \textbf{TREC DL} & \textbf{BEIR} (5 subsets) \\
 \midrule
 \textit{yes / no} & 20.8 & 60.9 & 31.1 \\
 $\{0, 1, \cdots, 3\}$ & 22.7 & 64.1 & 36.5 \\
 $\{0, 1, \cdots, 10\}$ & 23.2 & 66.1 & 37.1 \\
 \bottomrule
 \end{tabular}
 \caption{Average nDCG@10 of QwQ-32B reasoning LLM when varying scoring discrimination on three benchmarks.}
 \label{tab:discrimination}
\end{table}

Such a high-quality dataset $\mathcal{D}$ enables the model to learn nuanced relevance assessment. The model is then trained on this dataset using a standard language modeling objective:

{
\vspace{-2mm}
\begin{equation*}
\mathcal{L}_{SFT}(\theta) = - \sum_{(x, y) \in \mathcal{D}} \log P(y \mid x;\theta).
\end{equation*}
}

\subsection{Reinforcement Learning with GRPO}
\label{sec:RL}

While Supervised Fine-Tuning (SFT) provides a strong reranking model, we employ Reinforcement Learning (RL) to further refine its ability to discern subtle ranking differences and optimize for list level metrics. To achieve this, we adopt the Group Relative Policy Optimization (GRPO) algorithm~\cite{GRPO}, inspired by prior work demonstrating that RL on small, high quality datasets can yield significant performance gains~\cite{DeepSeek-R1}. We initialize both the GRPO policy $\pi_\theta$ and the reference model $\pi_\text{ref}$ with the SFT-tuned model to ensure training stability and preserve its well generalized capabilities.

The training process begins by sampling a group of $G$ output trajectories $\{y_1, y_2, \dots, y_G\}$ for each input prompt with the old policy $\pi_\text{old}$. The policy $\pi_\theta$ is then updated by optimizing the GRPO objective. This objective is built around a clipped importance sampling estimator, which evaluates the advantage of each trajectory relative to others in the group. To prevent the policy from deviating too drastically from the robust SFT model, we incorporate a Kullback Leibler (KL) divergence penalty. This term regularizes the policy updates, ensuring the model learns a more nuanced scoring function without sacrificing its foundational knowledge. The complete objective function is formulated as follows:

{\small
\vspace{-4mm}
\begin{equation*}
 J_{GRPO}(\theta) = \mathbb{E}_{x \sim \mathcal{D}, \{y_i\} \sim \pi_{\theta_{\text{old}}}} \left[ \frac{1}{G} \sum_{i=1}^{G} \frac{1}{|y_i|} \sum_{t=1}^{|y_i|} \left( \mathcal{C} - \beta D_\text{KL}
 \right) \right],
\end{equation*}
}
where the clipped estimator $\mathcal{C}$ and KL penalty $D_\text{KL}$ are:
{\small
\vspace{-1mm}
\begin{align*}
 \mathcal{C} = \min & ( \frac{\pi_\theta(y_{i,t}|x, y_{i,<t})}{\pi_{\theta_\text{old}}(y_{i,t}|x, y_{i,<t})} \hat{A}_{i,t}, \\
 & \text{clip} \left( \frac{\pi_\theta(y_{i,t}|x, y_{i,<t})}{\pi_{\theta_\text{old}}(y_{i,t}|x, y_{i,<t})}, 1-\epsilon, 1+\epsilon \right) \hat{A}_{i,t} ),
\end{align*}
\begin{equation*}
 D_\text{KL}
 = \frac{\pi_{\text{ref}}(y_{i,t}|x, y_{i,<t})}{\pi_\theta(y_{i,t}|x, y_{i,<t})} - \log \frac{\pi_{\text{ref}}(y_{i,t}|x, y_{i,<t})}{\pi_\theta(y_{i,t}|x, y_{i,<t})} - 1,
\end{equation*}
}where the advantage $\hat{A}_{i,t}$ is computed by normalizing the reward $r(y_i)$ of trajectory $y_i$ using the mean and standard deviation of rewards in the group: $\hat{A}_{i,t} = \frac{r(y_i) - \text{mean}(\{r(y_j)\}_{j=1}^G)}{\text{std}(\{r(y_j)\}_{j=1}^G)}$. The hyperparameters $\beta$ and $\epsilon$ control the strength of the KL penalty and the clipping range, respectively.

\begin{figure}[t]
 \centering
 \includegraphics[width=\linewidth]{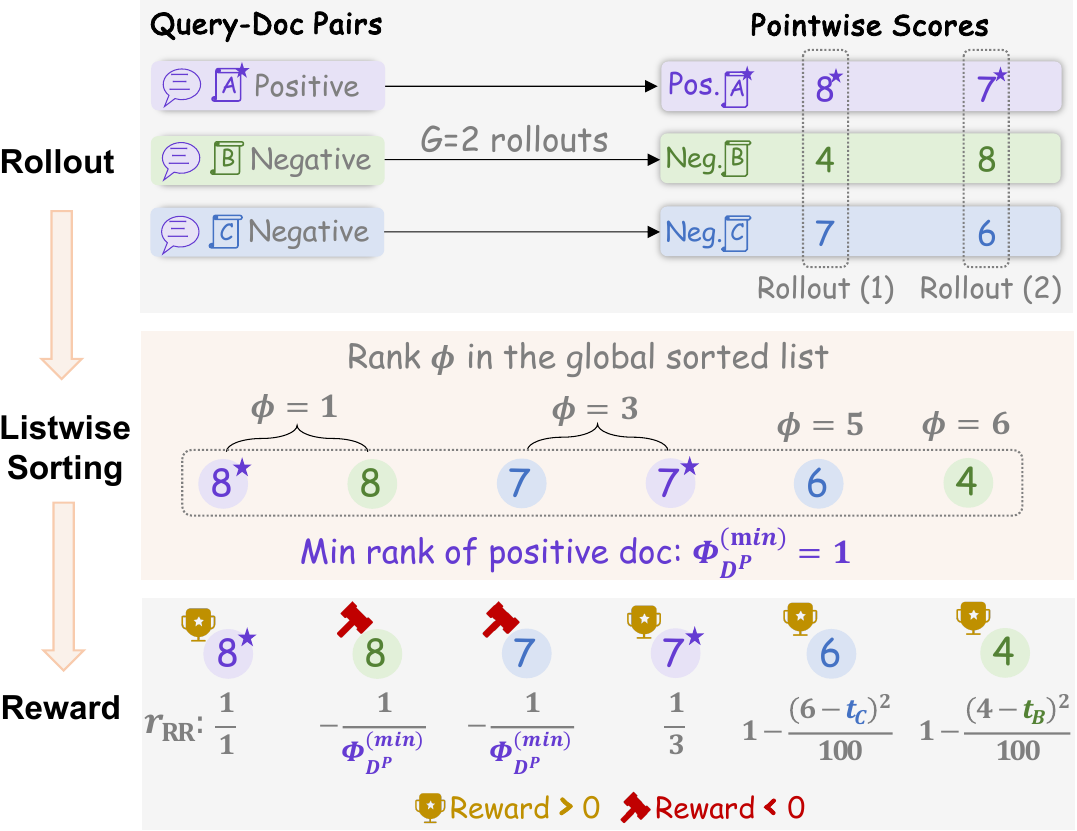}
 \caption{Example for rule-based listwise reward $r_\text{RR}$ when there are $G=2$ rollouts and $N=3$ documents for query $q$.}
 \label{fig:reward}
\end{figure}

\paragraph{Listwise Reranking Reward Design.}
\label{sec:listwise_reward}
Previous studies have established that listwise rerankers often outperform pointwise models because they directly optimize the relative ordering of documents. Drawing inspiration from this, we designed a novel rule-based listwise reward function, $r_{\text{RR}}$. While our model produces scores pointwise at inference, this reward design allows it to learn from relative document ordering during training, a key strength of listwise methods.

As shown in Figure~\ref{fig:reward}, for a given query, we first generate scores for all $N$ candidate documents and their corresponding $G$ rollouts. These $N \times G$ scores are then aggregated and sorted to determine the global rank of each generated output. Our reward function, $r_{\text{RR}}$, operates based on the following principles. Positive documents receive a high reward based on their reciprocal rank, incentivizing them to be placed as high as possible. Negative documents that are incorrectly ranked higher than any positive document receive a substantial penalty. Negative documents ranked correctly below all positive documents receive a smooth reward based on the squared error against a reference score from the SFT model, which helps maintain a stable scoring distribution. A significant penalty is assigned if the model's output is not correctly formatted, which discourages generation errors.

The formal definition of the reward $r_{\text{RR}}$ is as follows, where $\phi_i^{(j)}$ is the global rank of the $j$-th rollout for document $d_i$ with ties resolved by assigning the minimum rank.
Let $\mathcal{D}^P$ and $\mathcal{D}^N$ be the set of positive and negative documents, respectively.
Let $\Phi_{\mathcal{D}^P}^{(\text{min})}$ and $\Phi_{\mathcal{D}^P}^{(\text{max})}$ denote the minimum and maximum ranks for positive documents, respectively.

{\small
\vspace{-3mm}
\begin{equation*}
 r_\text{RR} =
 \begin{cases}
 \frac{1}{\phi_i^{(j)}}, & \text{formatted, } d_i \in \mathcal{D}^P, \\
 -\frac{1}{\Phi_{\mathcal{D}^P}^{\text{(min)}}}, & \text{formatted, } d_i \in \mathcal{D}^N, \phi_i^{(j)} \leq \Phi_{\mathcal{D}^P}^{\text{(max)}}, \\
 1 - \frac{(s_i - t_i)^2}{100}, & \text{formatted, } d_i \in \mathcal{D}^N, \phi_i^{(j)} > \Phi_{\mathcal{D}^P}^{\text{(max)}}, \\
 -1, & \text{otherwise},
 \end{cases}
\end{equation*}
}where ``formatted'' condition requires that the model's output conforms to the expected structure so a score can be extracted. In the third case, $s_i$ is the generated score and $t_i$ is a reference score from the SFT model $\pi_{\text{ref}}$.

We trained the model using a randomly sampled subset of $2,048$ queries from the SFT dataset, each paired with $20$ documents. The GRPO algorithm was applied directly using these pointwise prompts without any pre-collected trajectories or external reward signals. The SFT tuned model served a dual role as both the initial policy for training and the reference model $\pi_{\text{ref}}$ for calculating KL divergence and providing reference scores for the reward function.

\begin{table}[t]
 \centering
 \small
 \begin{tabular}{c|ccc}
 \toprule
 \textbf{Stage} & \textbf{\# Queries} & \textbf{\# Doc per query} & \textbf{\# Query-doc pairs} \\
 \midrule
 SFT & 14,799 & 20 & 295,980 \\
 RL & 2,048 & 20 & 40,960 \\
 \bottomrule
 \end{tabular}
 \caption{Statistics of training data.}
 \label{tab:training_data}
\end{table}

\begin{table*}[t]
\small
\centering
\setlength{\tabcolsep}{3.8mm}{
\begin{tabular}{llccccc}
\toprule
\multicolumn{2}{l}{\multirow{2}{*}{\textbf{Method}}} & \multirow{2}{*}{\textbf{Average}} & \multicolumn{2}{c}{\textbf{Reasoning-intensive Relevance}} & \multicolumn{2}{c}{\textbf{Semantic Relevance}} \\
\cmidrule(lr){4-5} \cmidrule(lr){6-7}
\multicolumn{2}{l}{} & & BRIGHT & FollowIR & BEIR & TREC DL \\
\midrule
\multicolumn{2}{l}{First-stage retriever} & 25.9 & 13.7 & 0 & 40.8 & 49.3 \\
\midrule
\multirow{2}{*}{Listwise} & Rank-R1-7B~\cite{Rank-R1} & 34.6 & 15.7 & 3.6 & \textbf{49.0} & \underline{70.0} \\
 & Rearank-7B~\cite{Rearank} & 35.3 & 17.4 & 2.3 & \textbf{49.0} & \textbf{72.5} \\
\midrule
\multirow{6}{*}{Pointwise} & JudgeRank-8B~\cite{JudgeRank} & 32.1 & 17.0 & 9.9 & 39.1 & 62.6 \\
 & Rank1-7B~\cite{Rank1} & 34.6 & 18.2 & 9.1 & 44.2 & 67.1 \\
 & QwQ-32B powered reranker \textbf{(Ours)} & \underline{37.7} & \underline{23.2} & \textbf{14.1} & 47.5 & 66.1 \\
 & {\methodname}-4B \textbf{(Ours)} & 36.8 & 22.7 & 11.0 & 44.8 & 68.9 \\
 & {\methodname}-14B \textbf{(Ours)} & 36.9 & 23.1 & 10.3 & 47.1 & 67.1 \\
 & {\methodname}-32B \textbf{(Ours)} & \textbf{38.1} & \textbf{24.4} & \underline{12.1} & \underline{47.7} & 68.1 \\
\bottomrule
\end{tabular}
}
\caption{Evaluation on different relevance types using original queries without hybrid scores. Best results are indicated in bold, while second-best results are underlined.}
\label{tab:main_result}
\end{table*}

\begin{table}[t]
\small
\centering
\begin{tabular}{p{61mm}l}
\toprule
\textbf{Method} & \textbf{nDCG@10} \\
\midrule
\multicolumn{2}{c}{\textit{Retrieve by BM25 using GPT-4 reason-query}} \\
\midrule
BM25 & 27.0 \\
Rank-R1-7B~\cite{Rank-R1} & 23.9 \\
Rank1-7B~\cite{Rank1} & 25.5 \\
Rearank-7B~\cite{Rearank} & 29.1 \\
XRR2-Gemini-2.5-Flash~\cite{XRR2} & \textbf{40.3}* \\
JudgeRank-8B~\cite{JudgeRank} & 24.4 \\
\textit{+ BM25 hybrid} & 31.0 \\
{\methodname}-4B \textbf{(Ours)} & 32.9 \\
\textit{+ BM25 hybrid} & 36.1 \\
{\methodname}-14B \textbf{(Ours)} & 33.5 \\
\textit{+ BM25 hybrid} & 36.7 \\
{\methodname}-32B \textbf{(Ours)} & 34.6 \\
\textit{+ BM25 hybrid} & \underline{37.4} \\
\midrule
\multicolumn{2}{c}{\textit{Retrieve by ReasonIR-8B using GPT-4 reason-query}} \\
\midrule
ReasonIR-8B~\cite{ReasonIR} & 30.5 \\
Rank-R1-7B~\cite{Rank-R1} & 24.1 \\
Rank1-7B~\cite{Rank1} & 24.3 \\
Rearank-7B~\cite{Rearank} & 27.5 \\
JudgeRank-8B~\cite{JudgeRank} & 20.2 \\
\textit{+ BM25 hybrid} & 22.7 \\
Rank-R1-32B-v0.2~\cite{Rank-R1-v0.2} & 37.7* \\
\textit{+ BM25 hybrid} & \underline{40.0}* \\
{\methodname}-4B \textbf{(Ours)} & 30.5 \\
\textit{+ BM25 hybrid} & 38.7 \\
{\methodname}-14B \textbf{(Ours)} & 31.8 \\
\textit{+ BM25 hybrid} & 39.3 \\
{\methodname}-32B \textbf{(Ours)} & 32.8 \\
\textit{+ BM25 hybrid} & \textbf{40.2} \\
\bottomrule
\end{tabular}
\caption{Further evaluation on BRIGHT benchmark. *Taken from BRIGHT online website \cite{BRIGHTWebsite}.}
\label{tab:bright_brief}
\end{table}

\section{Experiment}
\label{sec:exp}

\subsection{Evaluation Setup}

\paragraph{Benchmarks.}
We evaluate our method across a diverse set of reranking tasks on four benchmark suites. For in-domain semantic matching, we use the TREC DL19 and DL20 passage ranking collections~\cite{DL19, DL20}. For out-of-domain generalization, we utilize the entire BEIR benchmark~\cite{BEIR} and report results on a five-dataset subset, which we term BEIR-5 (ArguAna, DBPedia, FiQA, NFCorpus, and SCIDOCS), to facilitate efficient ablation studies. To assess complex reasoning, we employ the BRIGHT benchmark~\cite{BRIGHT} for general reasoning and the FollowIR benchmark~\cite{FollowIR} for instruction following abilities.

\paragraph{Baselines.}
We compare our model, {\methodname}, against leading reasoning-based rerankers. These include JudgeRank-8B, a zero-shot reranker that uses a multi-step reasoning process; Rank1-7B, a pointwise reranker fine-tuned via distillation; Rank-R1-7B, a listwise reranker trained with the GRPO algorithm; and Rearank-7B, a state-of-the-art listwise model trained to predict optimal document permutations. Additionally, we include two top-performing listwise rerankers from the online BRIGHT benchmark~\cite{BRIGHTWebsite}: Rank-R1-32B-v0.2~\cite{Rank-R1-v0.2} trained on the ReasonIR training set, and the zero-shot XRR-Gemini-2.5-Flash~\cite{XRR2} which performs a two-pass reranking process.

\subsection{Implementation and Evaluation Details}
\label{sec:implementation}

For TREC DL, BEIR, and BRIGHT benchmarks, we rerank the top $100$ candidates retrieved by BM25 with Pyserini~\cite{pyserini}. For FollowIR, we rerank the 1,000 candidates provided by the benchmark. Evaluation is performed using nDCG@10 for the first three benchmarks and preference-based Mean Reciprocal Rank ($p$-MRR) for FollowIR.  In all cases, higher scores indicate better performance.

Following prior studies~\cite{Rank1, JudgeRank, Rank-R1-v0.2}, we adopt similar settings for a thorough evaluation on the challenging BRIGHT benchmark. First, to improve first-stage retrieval precision, we use reasoning queries expanded by GPT-4. Documents are then retrieved using either BM25 or the ReasonIR-8B model~\cite{ReasonIR}. During the reranking phase, these expanded queries are not provided to the reranker. Second, we employ a hybrid strategy to combine BM25 scores and reranking scores for low-cost model ensembling. While JudgeRank uses a simple weighted sum and Rank-R1-32B-v0.2 adopts min-max normalization, our {\methodname} applies standardization before score aggregation, as detailed in Appendix~\ref{app:bright_settings}.

Using Qwen3 LLM series~\cite{Qwen3} as the backbone, our {\methodname} model is trained in two-stages. The first stage consists of one epoch of Supervised Fine-Tuning (SFT) with Low-Rank Adaptation (LoRA)~\cite{lora}. The second stage uses the GRPO algorithm~\cite{GRPO} for Reinforcement Learning (RL), performing full-parameter fine-tuning for $10$ epochs with a group $G=5$. Detailed hyperparameters can be found in Appendices~\ref{app:SFT} and~\ref{app:RL}. As shown in Table~\ref{tab:training_data}, after filtering out those longer than $2,048$ response tokens, there are $14,799$ queries and $2,048$ queries for SFT and RL training, respectively. All experiments are conducted on four NVIDIA A100 (80GB) GPUs. We use official checkpoints for all baselines and reproduce JudgeRank based on its published methodology.

\subsection{Main Results}
\label{sec:main_results}

Table~\ref{tab:main_result} presents the main results, with detailed reports available in Appendix~\ref{app:detailed_results}. On average, {\methodname}-4B clearly outperforms all pointwise baselines with 7B or 8B parameters. Furthermore, {\methodname}-4B significantly surpasses listwise rerankers, which are typically more powerful, on reasoning-intensive tasks despite having fewer parameters. This demonstrates that {\methodname}-4B achieves superior effectiveness while maintaining lower latency compared to listwise methods. Beyond the 4B model, we extend our two-stage training pipeline to Qwen3-14B and Qwen3-32B models using identical training data. The results show an overall performance improvement with increased model size, indicating a clear scaling trend. At the 32B scale, our trained {\methodname}-32B reranker outperforms its teacher model, QwQ-32B, which confirms the efficacy of our training procedure.

Table~\ref{tab:bright_brief} further reports results with advanced retrieval and BM25 hybrid strategy on the BRIGHT benchmark. Our {\methodname} rerankers consistently achieve state-of-the-art performance compared to baselines of similar model size, showing superior robustness and effectiveness. Despite using a pointwise paradigm, {\methodname}-4B achieves a notable nDCG@10 of 38.7 with the BM25 hybrid on documents retrieved by ReasonIR-8B. The {\methodname}-32B model with BM25 hybrid achieves an nDCG@10 of 40.2, outperforming the state-of-the-art Rank-R1-32B-v0.2 listwise reranker. Moreover, {\methodname}-32B approaches the performance of XRR2, a listwise method that employs the Gemini-2.5-Flash model.

\subsection{Analysis}
\label{sec:ablation}

\begin{table}[t]
\small
\centering
\setlength{\tabcolsep}{1.4mm}{
\begin{tabular}{l|c|cccc}
\toprule
 & \textbf{Avg.} & BRIGHT & FollowIR & BEIR-5 & TREC DL \\
 \midrule
Qwen3-4B & 12.7 & 3.6 & 1.9 & 6.4 & 39.0 \\
SFT Only & \underline{32.8} & \underline{22.0} & \underline{11.2} & \underline{30.0} & \underline{68.1} \\
RL Only & 31.8 & 20.1 & \textbf{12.2} & 28.2 & 66.5 \\
SFT+RL & \textbf{33.8} & \textbf{22.7} & 11.0 & \textbf{32.4} & \textbf{68.9} \\
\bottomrule
\end{tabular}
}
\caption{Performance of different training stages.}
\label{tab:train_stage}
\end{table}

\begin{table}[t]
\small
\centering
\setlength{\tabcolsep}{1.44mm}{
\begin{tabular}{l|c|cccc}
\toprule
 & \textbf{Avg.} & BRIGHT & FollowIR & BEIR-5 & TREC DL \\
\midrule
Before RL & \underline{32.8} & 22.0 & \underline{11.2} & 30.0 & \underline{68.1} \\
$r_\text{SE}$ & 31.1 & \underline{22.3} & 8.7 & 30.8 & 62.6 \\
$r_\text{nDCG}$ & \textbf{33.8} & 21.5 & \textbf{13.2} & \textbf{32.8} & 67.6 \\
$r_\text{RR}$ & \textbf{33.8} & \textbf{22.7} & 11.0 & \underline{32.4} & \textbf{68.9} \\
\bottomrule 
\end{tabular}
}
\caption{Performance of different rewards for RL training.}
\label{tab:reward}
\end{table}

\paragraph{Impact of training stages.}
To investigate the contribution of SFT and RL to reranking ability, we perform an ablation study using consistent prompts across different model variants. As shown in Table~\ref{tab:train_stage}, the instructed Qwen3-4B LLM without any fine-tuning performs poorly. Both SFT and RL independently yield significant improvements, highlighting their individual effectiveness. Our two-stage training pipeline for {\methodname}-4B yields the most robust effectiveness overall.

\paragraph{Varying rewards in RL training.}
Besides the rule-based listwise reward $r_\text{RR}$ using Reciprocal Rank, we evaluate two different rule-based rewards, which are briefly described as follows. Please refer to Appendix~\ref{app:rewards} for more details.
\begin{itemize}[leftmargin=*]
 \item \textit{Pointwise reward $r_\text{SE}$.} It uses squared error to measure the difference between the score $s_i$ from the policy model and the score $t_i$ from the teacher model (i.e., QwQ-32B).
 \item \textit{Listwise reward $r_\text{nDCG}$.} This is a listwise reward similar to $r_\text{RR}$, which assesses how effectively positive documents contribute to the nDCG metric while penalizing negative documents ranked above any positive document.
\end{itemize}

\begin{figure}[!htp]
 \centering
 \includegraphics[width=\linewidth]{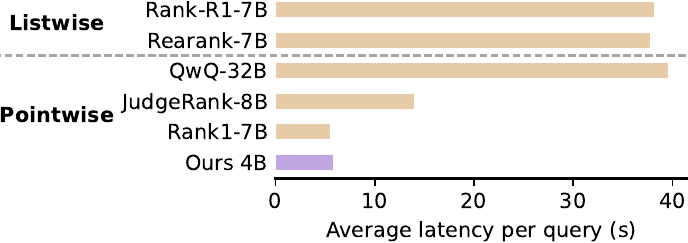}
 \caption{Latency for returning the complete reranked list per query, averaged on all queries of TREC DL19 dataset.}
 \label{fig:latency}
\end{figure}

Table~\ref{tab:reward} compares the results when utilizing different rewards for GRPO training. Overall, listwise rewards such as $r_\text{nDCG}$ and $r_\text{RR}$ lead to better outcomes than the pointwise reward $r_\text{SE}$. Pointwise reward that mimicks the teacher model's scores for each document independently may not align well with global ranking objectives. In contrast, listwise rewards tend to yield more favorable results by considering relative ranks to encourage a better final reranking order. While $r_\text{nDCG}$ shows a notable improvement on the FollowIR benchmark, $r_\text{RR}$ demonstrates greater robustness and superior overall performance across the four benchmarks.

\paragraph{Reranking latency.}
In real-world applications, achieving superior performance across diverse relevance types must be balanced with acceptable latency. As shown in Figure~\ref{fig:latency}, which measures latency per query on the TREC DL19 dataset, pointwise methods offer significantly lower latency than their listwise counterparts. This advantage stems from the ability of pointwise methods to process documents in parallel, whereas listwise methods require sequential processing as discussed in Section~\ref{sec:related_work}. Specifically, the {\methodname}-4B reranker is six times faster than both the listwise methods and the QwQ-32B pointwise reranker, highlighting its practicality for real-world applications.
Compared to Rank1-7B, {\methodname}-4B achieves superior performance by generating more tokens while maintaining comparable latency.

\section{Conclusion}
\label{sec:conclusion}
In this paper, we introduce {\methodname}, an LLM-based reranker designed for effective and efficient reranking of documents in both semantic and reasoning-intensive tasks. To support real-world applications, {\methodname} adopts the pointwise paradigm to ensure low latency while achieving competitive performance through a two-stage training pipeline. The first stage conducts Supervised Fine-Tuning (SFT) to build foundational reasoning capabilities, and the second stage employs the GRPO algorithm with a novel rule-based listwise reward tailored for pointwise rerankers. Extensive evaluation on four benchmarks demonstrates the effectiveness and robustness of {\methodname} compared to state-of-the-art methods.

\bibliography{citation}

\begin{thebibliography}{45}
\providecommand{\natexlab}[1]{#1}

\bibitem[{Bajaj et~al.(2016)Bajaj, Campos, Craswell, Deng, Gao, Liu, Majumder, McNamara, Mitra, Nguyen et~al.}]{MS-MARCO}
Bajaj, P.; Campos, D.; Craswell, N.; Deng, L.; Gao, J.; Liu, X.; Majumder, R.; McNamara, A.; Mitra, B.; Nguyen, T.; et~al. 2016.
\newblock {MS MARCO}: A human generated machine reading comprehension dataset.
\newblock \emph{arXiv preprint arXiv:1611.09268}.

\bibitem[{Chuang et~al.(2020)Chuang, Chen, Wang, Tsai, Fang, and Lim}]{chuang2020tpr}
Chuang, Y.-N.; Chen, C.-M.; Wang, C.-J.; Tsai, M.-F.; Fang, Y.; and Lim, E.-P. 2020.
\newblock TPR: Text-aware preference ranking for recommender systems.
\newblock In \emph{Proceedings of the 29th ACM International Conference on Information \& Knowledge Management}, 215--224.

\bibitem[{Craswell et~al.(2021)Craswell, Mitra, Yilmaz, and Campos}]{DL20}
Craswell, N.; Mitra, B.; Yilmaz, E.; and Campos, D. 2021.
\newblock Overview of the TREC 2020 deep learning track.
\newblock arXiv:2102.07662.

\bibitem[{Craswell et~al.(2020)Craswell, Mitra, Yilmaz, Campos, and Voorhees}]{DL19}
Craswell, N.; Mitra, B.; Yilmaz, E.; Campos, D.; and Voorhees, E.~M. 2020.
\newblock Overview of the TREC 2019 deep learning track.
\newblock arXiv:2003.07820.

\bibitem[{{DeepSeek AI}(2025)}]{DeepSeek-R1}
{DeepSeek AI}. 2025.
\newblock DeepSeek-R1: Incentivizing Reasoning Capability in LLMs via Reinforcement Learning.
\newblock arXiv:2501.12948.

\bibitem[{Gao et~al.(2025)Gao, Chen, Zhao, Liu, Li, Wang, Wang, Guo, and Tang}]{gao2025llm4rerank}
Gao, J.; Chen, B.; Zhao, X.; Liu, W.; Li, X.; Wang, Y.; Wang, W.; Guo, H.; and Tang, R. 2025.
\newblock Llm4rerank: Llm-based auto-reranking framework for recommendations.
\newblock In \emph{Proceedings of the ACM on Web Conference 2025}, 228--239.

\bibitem[{Gao, Dai, and Callan(2021)}]{rethinkBERT}
Gao, L.; Dai, Z.; and Callan, J. 2021.
\newblock Rethink training of BERT rerankers in multi-stage retrieval pipeline.
\newblock In \emph{European Conference on Information Retrieval}, 280--286. Springer.

\bibitem[{Gupta, Ranjan, and Singh(2024)}]{RAG-survey-2}
Gupta, S.; Ranjan, R.; and Singh, S.~N. 2024.
\newblock A comprehensive survey of retrieval-augmented generation {(RAG)}: Evolution, current landscape and future directions.
\newblock \emph{arXiv preprint arXiv:2410.12837}.

\bibitem[{Hu et~al.(2022)Hu, Shen, Wallis, Allen-Zhu, Li, Wang, Wang, Chen et~al.}]{lora}
Hu, E.~J.; Shen, Y.; Wallis, P.; Allen-Zhu, Z.; Li, Y.; Wang, S.; Wang, L.; Chen, W.; et~al. 2022.
\newblock Lora: Low-rank adaptation of large language models.
\newblock \emph{ICLR}, 1(2): 3.

\bibitem[{Huang et~al.(2024)Huang, Liu, Chen, Wang, Wang, Lian, Wang, Tang, and Chen}]{huang2024understanding}
Huang, X.; Liu, W.; Chen, X.; Wang, X.; Wang, H.; Lian, D.; Wang, Y.; Tang, R.; and Chen, E. 2024.
\newblock Understanding the planning of LLM agents: A survey.
\newblock \emph{arXiv preprint arXiv:2402.02716}.

\bibitem[{ielabgroup(2025)}]{Rank-R1-v0.2}
ielabgroup. 2025.
\newblock {Rank-R1-32B-v0.2}.
\newblock \url{https://huggingface.co/ielabgroup/Rank-R1-32B-v0.2}.
\newblock Accessed: 2025-07-24.

\bibitem[{jataware(2025)}]{XRR2}
jataware. 2025.
\newblock XRR2: Expand $\rightarrow$ Retrieve $\rightarrow$ Rerank $\rightarrow$ Rerank - simple method with strong results on BRIGHT benchmark.
\newblock \url{https://github.com/jataware/XRR2}.
\newblock Accessed: 2025-07-24.

\bibitem[{Lee et~al.(2018)Lee, Yun, Kim, Ko, and Kang}]{lee2018ranking}
Lee, J.; Yun, S.; Kim, H.; Ko, M.; and Kang, J. 2018.
\newblock Ranking Paragraphs for Improving Answer Recall in Open-Domain Question Answering.
\newblock In \emph{Proceedings of the 2018 Conference on Empirical Methods in Natural Language Processing}, 565--569.

\bibitem[{Li et~al.(2024)Li, Wang, Zeng, Wu, and Yang}]{li2024survey}
Li, X.; Wang, S.; Zeng, S.; Wu, Y.; and Yang, Y. 2024.
\newblock A survey on LLM-based multi-agent systems: workflow, infrastructure, and challenges.
\newblock \emph{Vicinagearth}, 1(1): 9.

\bibitem[{Liang et~al.(2022)Liang, Bommasani, Lee, Tsipras, Soylu, Yasunaga, Zhang, Narayanan, Wu, Kumar et~al.}]{liang2022holistic}
Liang, P.; Bommasani, R.; Lee, T.; Tsipras, D.; Soylu, D.; Yasunaga, M.; Zhang, Y.; Narayanan, D.; Wu, Y.; Kumar, A.; et~al. 2022.
\newblock Holistic evaluation of language models.
\newblock \emph{arXiv preprint arXiv:2211.09110}.

\bibitem[{Lin et~al.(2021)Lin, Ma, Lin, Yang, Pradeep, and Nogueira}]{pyserini}
Lin, J.; Ma, X.; Lin, S.-C.; Yang, J.-H.; Pradeep, R.; and Nogueira, R. 2021.
\newblock Pyserini: A Python toolkit for reproducible information retrieval research with sparse and dense representations.
\newblock In \emph{Proceedings of the 44th International ACM SIGIR Conference on Research and Development in Information Retrieval}, 2356--2362.

\bibitem[{Lin, Nogueira, and Yates(2022)}]{lin2022pretrained}
Lin, J.; Nogueira, R.; and Yates, A. 2022.
\newblock \emph{Pretrained transformers for text ranking: Bert and beyond}.
\newblock Springer Nature.

\bibitem[{Liu et~al.(2025)Liu, Ma, Zhao, Zheng, Ma, and Kang}]{liu2025listconranker}
Liu, J.; Ma, Y.; Zhao, R.; Zheng, J.; Ma, Q.; and Kang, Y. 2025.
\newblock ListConRanker: A Contrastive Text Reranker with Listwise Encoding.
\newblock \emph{arXiv preprint arXiv:2501.07111}.

\bibitem[{Liu et~al.(2024)Liu, Zhou, Zhu, Lian, Li, Dou, Lian, and Nie}]{liu2024information}
Liu, Z.; Zhou, Y.; Zhu, Y.; Lian, J.; Li, C.; Dou, Z.; Lian, D.; and Nie, J.-Y. 2024.
\newblock Information retrieval meets large language models.
\newblock In \emph{Companion Proceedings of the ACM Web Conference 2024}, 1586--1589.

\bibitem[{Ma et~al.(2024)Ma, Wang, Yang, Wei, and Lin}]{repLLaMA}
Ma, X.; Wang, L.; Yang, N.; Wei, F.; and Lin, J. 2024.
\newblock Fine-tuning llama for multi-stage text retrieval.
\newblock In \emph{Proceedings of the 47th International ACM SIGIR Conference on Research and Development in Information Retrieval}, 2421--2425.

\bibitem[{Ma et~al.(2023)Ma, Zhang, Pradeep, and Lin}]{ma2023zero}
Ma, X.; Zhang, X.; Pradeep, R.; and Lin, J. 2023.
\newblock Zero-shot listwise document reranking with a large language model.
\newblock \emph{arXiv preprint arXiv:2305.02156}.

\bibitem[{Niu et~al.(2024)Niu, Joty, Liu, Xiong, Zhou, and Yavuz}]{JudgeRank}
Niu, T.; Joty, S.; Liu, Y.; Xiong, C.; Zhou, Y.; and Yavuz, S. 2024.
\newblock JudgeRank: Leveraging Large Language Models for Reasoning-Intensive Reranking.
\newblock \emph{arXiv preprint arXiv:2411.00142}.

\bibitem[{OpenAI(2024)}]{OpenAI-O1}
OpenAI. 2024.
\newblock OpenAI o1 System Card.
\newblock arXiv:2412.16720.

\bibitem[{Qin et~al.(2024)Qin, Jagerman, Hui, Zhuang, Wu, Yan, Shen, Liu, Liu, Metzler et~al.}]{qin2024large}
Qin, Z.; Jagerman, R.; Hui, K.; Zhuang, H.; Wu, J.; Yan, L.; Shen, J.; Liu, T.; Liu, J.; Metzler, D.; et~al. 2024.
\newblock Large Language Models are Effective Text Rankers with Pairwise Ranking Prompting.
\newblock In \emph{Findings of the Association for Computational Linguistics: NAACL 2024}, 1504--1518.

\bibitem[{{Qwen Team}(2025{\natexlab{a}})}]{Qwen3}
{Qwen Team}. 2025{\natexlab{a}}.
\newblock Qwen3 Technical Report.
\newblock arXiv:2505.09388.

\bibitem[{{Qwen Team}(2025{\natexlab{b}})}]{QwQ}
{Qwen Team}. 2025{\natexlab{b}}.
\newblock {QwQ}-32B: Embracing the Power of Reinforcement Learning.

\bibitem[{Shao et~al.(2025)Shao, Qiao, Kishore, Muennighoff, Lin, Rus, Low, Min, Yih, Koh et~al.}]{ReasonIR}
Shao, R.; Qiao, R.; Kishore, V.; Muennighoff, N.; Lin, X.~V.; Rus, D.; Low, B. K.~H.; Min, S.; Yih, W.-t.; Koh, P.~W.; et~al. 2025.
\newblock {ReasonIR}: Training Retrievers for Reasoning Tasks.
\newblock \emph{arXiv preprint arXiv:2504.20595}.

\bibitem[{Shao et~al.(2024)Shao, Wang, Zhu, Xu, Song, Bi, Zhang, Zhang, Li, Wu et~al.}]{GRPO}
Shao, Z.; Wang, P.; Zhu, Q.; Xu, R.; Song, J.; Bi, X.; Zhang, H.; Zhang, M.; Li, Y.; Wu, Y.; et~al. 2024.
\newblock {DeepSeekMath}: Pushing the limits of mathematical reasoning in open language models.
\newblock \emph{arXiv preprint arXiv:2402.03300}.

\bibitem[{Sheng et~al.(2024)Sheng, Zhang, Ye, Wu, Zhang, Zhang, Peng, Lin, and Wu}]{verl}
Sheng, G.; Zhang, C.; Ye, Z.; Wu, X.; Zhang, W.; Zhang, R.; Peng, Y.; Lin, H.; and Wu, C. 2024.
\newblock {HybridFlow}: A Flexible and Efficient RLHF Framework.
\newblock \emph{arXiv preprint arXiv: 2409.19256}.

\bibitem[{Su et~al.(2024)Su, Yen, Xia, Shi, Muennighoff, Wang, Liu, Shi, Siegel, Tang, Sun, Yoon, Arik, Chen, and Yu}]{BRIGHT}
Su, H.; Yen, H.; Xia, M.; Shi, W.; Muennighoff, N.; Wang, H.-y.; Liu, H.; Shi, Q.; Siegel, Z.~S.; Tang, M.; Sun, R.; Yoon, J.; Arik, S.~O.; Chen, D.; and Yu, T. 2024.
\newblock {BRIGHT}: A Realistic and Challenging Benchmark for Reasoning-Intensive Retrieval.

\bibitem[{Su et~al.(2025)Su, Yen, Xia, Shi, Muennighoff, Wang, Liu, Shi, Siegel, Tang, Sun, Yoon, Arik, Chen, and Yu}]{BRIGHTWebsite}
Su, H.; Yen, H.; Xia, M.; Shi, W.; Muennighoff, N.; Wang, H.-y.; Liu, H.; Shi, Q.; Siegel, Z.~S.; Tang, M.; Sun, R.; Yoon, J.; Arik, S.~O.; Chen, D.; and Yu, T. 2025.
\newblock {BRIGHT Benchmark Online Website}.
\newblock \url{https://brightbenchmark.github.io/}.
\newblock Accessed: August 26, 2025.

\bibitem[{Sun et~al.(2023)Sun, Yan, Ma, Wang, Ren, Chen, Yin, and Ren}]{RankGPT}
Sun, W.; Yan, L.; Ma, X.; Wang, S.; Ren, P.; Chen, Z.; Yin, D.; and Ren, Z. 2023.
\newblock Is ChatGPT good at search? investigating large language models as re-ranking agents.
\newblock \emph{arXiv preprint arXiv:2304.09542}.

\bibitem[{Thakur et~al.(2021)Thakur, Reimers, R{\"u}ckl{\'e}, Srivastava, and Gurevych}]{BEIR}
Thakur, N.; Reimers, N.; R{\"u}ckl{\'e}, A.; Srivastava, A.; and Gurevych, I. 2021.
\newblock {BEIR}: A Heterogeneous Benchmark for Zero-shot Evaluation of Information Retrieval Models.
\newblock In \emph{Thirty-fifth Conference on Neural Information Processing Systems Datasets and Benchmarks Track (Round 2)}.

\bibitem[{Wang et~al.(2024)Wang, Wang, Gao, Zhang, Wu, Xu, Shi, Wang, Li, Qian et~al.}]{RAG-survey-3}
Wang, X.; Wang, Z.; Gao, X.; Zhang, F.; Wu, Y.; Xu, Z.; Shi, T.; Wang, Z.; Li, S.; Qian, Q.; et~al. 2024.
\newblock Searching for best practices in retrieval-augmented generation.
\newblock \emph{arXiv preprint arXiv:2407.01219}.

\bibitem[{Weller et~al.(2025{\natexlab{a}})Weller, Chang, MacAvaney, Lo, Cohan, Van~Durme, Lawrie, and Soldaini}]{FollowIR}
Weller, O.; Chang, B.; MacAvaney, S.; Lo, K.; Cohan, A.; Van~Durme, B.; Lawrie, D.; and Soldaini, L. 2025{\natexlab{a}}.
\newblock {F}ollow{IR}: Evaluating and Teaching Information Retrieval Models to Follow Instructions.
\newblock In Chiruzzo, L.; Ritter, A.; and Wang, L., eds., \emph{Proceedings of the 2025 Conference of the Nations of the Americas Chapter of the Association for Computational Linguistics: Human Language Technologies (Volume 1: Long Papers)}, 11926--11942. Albuquerque, New Mexico: Association for Computational Linguistics.

\bibitem[{Weller et~al.(2025{\natexlab{b}})Weller, Ricci, Yang, Yates, Lawrie, and Van~Durme}]{Rank1}
Weller, O.; Ricci, K.; Yang, E.; Yates, A.; Lawrie, D.; and Van~Durme, B. 2025{\natexlab{b}}.
\newblock Rank1: Test-time compute for reranking in information retrieval.
\newblock \emph{arXiv preprint arXiv:2502.18418}.

\bibitem[{Weller et~al.(2024)Weller, Van~Durme, Lawrie, Paranjape, Zhang, and Hessel}]{promptriever}
Weller, O.; Van~Durme, B.; Lawrie, D.; Paranjape, A.; Zhang, Y.; and Hessel, J. 2024.
\newblock Promptriever: Instruction-trained retrievers can be prompted like language models.
\newblock \emph{arXiv preprint arXiv:2409.11136}.

\bibitem[{Wu et~al.(2024)Wu, Xiong, Cui, Wu, Chen, Yuan, Huang, Liu, Kuo, Guan et~al.}]{RAG-survey-1}
Wu, S.; Xiong, Y.; Cui, Y.; Wu, H.; Chen, C.; Yuan, Y.; Huang, L.; Liu, X.; Kuo, T.-W.; Guan, N.; et~al. 2024.
\newblock Retrieval-augmented generation for natural language processing: A survey.
\newblock \emph{arXiv preprint arXiv:2407.13193}.

\bibitem[{Zhang et~al.(2023)Zhang, Chen, Liu, Niu, and Wang}]{zhang2023empirical}
Zhang, J.; Chen, Y.; Liu, C.; Niu, N.; and Wang, Y. 2023.
\newblock Empirical evaluation of ChatGPT on requirements information retrieval under zero-shot setting.
\newblock In \emph{2023 International Conference on Intelligent Computing and Next Generation Networks (ICNGN)}, 1--6. IEEE.

\bibitem[{Zhang et~al.(2025)Zhang, Wang, Qiu, Reddy, and Agrawal}]{Rearank}
Zhang, L.; Wang, B.; Qiu, X.; Reddy, S.; and Agrawal, A. 2025.
\newblock Rerank: Reasoning Re-ranking Agent via Reinforcement Learning.
\newblock \emph{arXiv preprint arXiv:2505.20046}.

\bibitem[{Zhang et~al.(2024)Zhang, Zhang, Long, Xie, Zhang, and Zhang}]{zhang2024two}
Zhang, L.; Zhang, Y.; Long, D.; Xie, P.; Zhang, M.; and Zhang, M. 2024.
\newblock A Two-Stage Adaptation of Large Language Models for Text Ranking.
\newblock In \emph{ACL (Findings)}.

\bibitem[{Zhang et~al.(2022)Zhang, Long, Xu, and Xie}]{zhang2022hlatr}
Zhang, Y.; Long, D.; Xu, G.; and Xie, P. 2022.
\newblock HLATR: enhance multi-stage text retrieval with hybrid list aware transformer reranking.
\newblock \emph{arXiv preprint arXiv:2205.10569}.

\bibitem[{Zheng et~al.(2024)Zheng, Zhang, Zhang, Ye, Luo, Feng, and Ma}]{llamafactory}
Zheng, Y.; Zhang, R.; Zhang, J.; Ye, Y.; Luo, Z.; Feng, Z.; and Ma, Y. 2024.
\newblock LlamaFactory: Unified Efficient Fine-Tuning of 100+ Language Models.
\newblock In \emph{Proceedings of the 62nd Annual Meeting of the Association for Computational Linguistics (Volume 3: System Demonstrations)}. Bangkok, Thailand: Association for Computational Linguistics.

\bibitem[{Zhuang et~al.(2025)Zhuang, Ma, Koopman, Lin, and Zuccon}]{Rank-R1}
Zhuang, S.; Ma, X.; Koopman, B.; Lin, J.; and Zuccon, G. 2025.
\newblock {Rank-R1}: Enhancing reasoning in llm-based document rerankers via reinforcement learning.
\newblock \emph{arXiv preprint arXiv:2503.06034}.

\bibitem[{Zhuang et~al.(2024)Zhuang, Zhuang, Koopman, and Zuccon}]{Setwise}
Zhuang, S.; Zhuang, H.; Koopman, B.; and Zuccon, G. 2024.
\newblock A setwise approach for effective and highly efficient zero-shot ranking with large language models.
\newblock In \emph{Proceedings of the 47th International ACM SIGIR Conference on Research and Development in Information Retrieval}, 38--47.

\end{thebibliography}

\begin{figure*}[t]
 \centering
 \includegraphics[width=\linewidth]{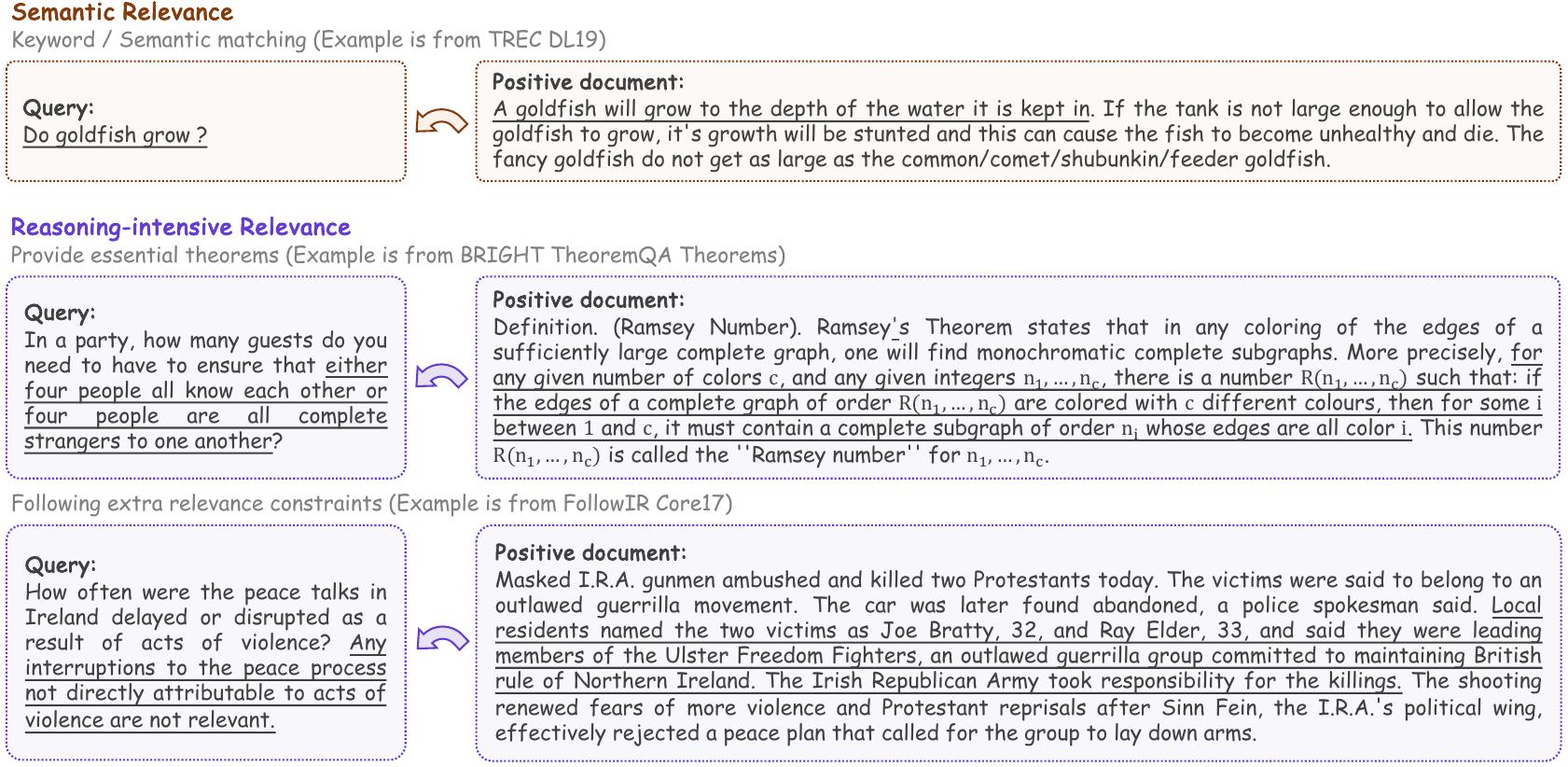}
 \caption{Illustrating examples for semantic relevance and reasoning-intensive relevance.}
 \label{fig:relevance}
\end{figure*}

\newpage
\appendix
\setcounter{secnumdepth}{1} 

\section{Relevance Types}
\label{app:relevance}

With examples in Figure~\ref{fig:relevance}, we introduce semantic and reasoning-intensive relevance as follows.

\paragraph{Semantic relevance.} 
It refers to the traditional understanding of relevance based on keyword or semantic matching between a query and a document~\cite{BEIR, DL19, DL20}. 
For example, the query \textit{``Do goldfish grow?''} can be lexically and semantically matched with \textit{``A goldfish will grow to the depth of the water it is kept in.''} in the positive document.

\paragraph{Reasoning-intensive relevance.}
Different from traditional semantic relevance, reasoning-intensive reranking should be able to capture documents that may not directly answer the query but provide essential intermediate information needed for multi-step reasoning \cite{BRIGHT}.
For example, the query requires that \textit{``either four people all know each other or four people are all complete strangers to one another''} cannot be directly answered by lexical or semantic matching with the positive document.
Instead, the document provides the essential mathematical foundation of \textit{Ramsey number} $R(n_1, \dots, n_c)$, which characterizes the minimum size of a complete graph such that any $c$-coloring of its edges contains a monochromatic complete subgraph of order $n_i$. 
The original query can be transformed into a case of \textit{Ramsey number}: guests at a party correspond to vertices, mutual acquaintance or stranger status corresponds to a $2$-color edge coloring, and the desired group of $4$ mutual acquaintances or strangers corresponds to a monochromatic $K_4$. Thus, solving the query reduces to determining $R(4,4)$. While the answer is not stated explicitly, the document supplies the critical intermediate concept required for multi-step reasoning.
Furthermore, queries with external constraints are also reasoning-intensive, determining which types of documents should or should not be considered relevant~\cite{FollowIR}.
For example, the query is about the disrupted peace in Ireland and requires that \textit{``Any interruptions to the peace process not directly attributable to acts of violence are not relevant.''}
Here, the positive document is about the violent conflict between two victims from a Northern Ireland outlawed guerrilla group and the Irish Republican Army.

\section{Preliminary Experiments}
\label{app:preliminary_exp}

We conduct preliminary experiments on BRIGHT~\cite{BRIGHT}, TREC DL~\cite{DL19, DL20}, and BEIR-5~\cite{BEIR} benchmarks in Section~\ref{sec:SFT}.
We use the original queries to retrieve top $100$ candidate documents using the pyserini implementation of BM25~\cite{pyserini}.
Then, we use these original queries and retrieved documents for rereanking under different settings as follows.

\begin{figure}[t]
 \centering
 \subfloat[TREC DL benchmark]{\includegraphics[width=\linewidth]{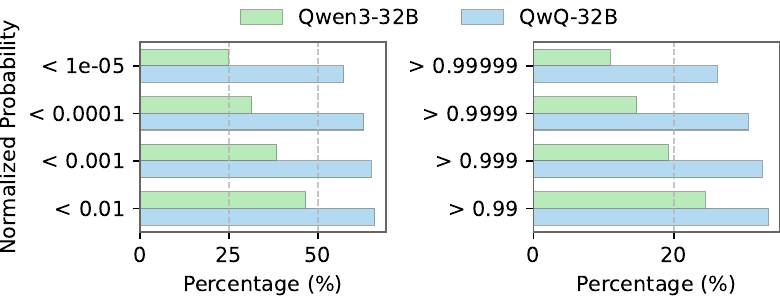}}\\
 \subfloat[BEIR benchmark (5 subsets)]{\includegraphics[width=\linewidth]{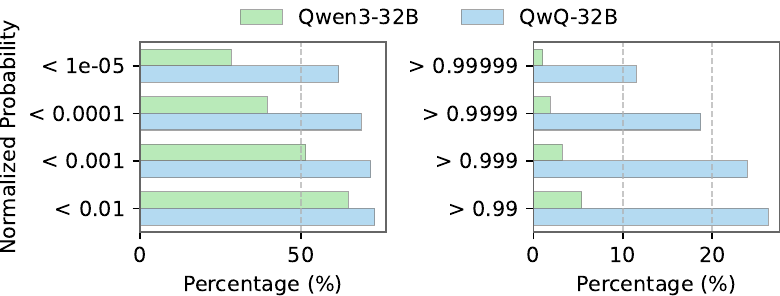}}
 \caption{Distributions of normalized probability with non-reasoning and reasoning LLMs on TREC DL and BEIR benchmarks.}
 \label{fig:probs_more}
\end{figure}

\paragraph{Comparing non-reasoning and reasoning LLMs.}
For non-reasoning LLM, we use Qwen3-32B and enable its non-reasoning mode to directly output a single word of \textit{``yes''} or \textit{``no''}.
For reasoning LLM, we use QwQ-32B~\cite{QwQ} that outputs Chain-of-Thought (CoT) before giving the binary judgement.
The corresponding prompts can be found in Appendix~\ref{app:prompts}.
After generation, we extract the probability of tokens \textit{``yes''} and \textit{``no''} to calculate the normalized probability as discussed in Section~\ref{sec:SFT}.
For each benchmark, we collect the normalized probabilities for all query-document pairs from all subsets to compute the ratios, after which we visualize them in Figures~\ref{fig:probs} and \ref{fig:probs_more}.

\paragraph{Comparing scoring discrimination.}
We use the same settings as above, except for using QwQ-32B~\cite{QwQ} and varying the scoring discrimination.
Specifically, we try scoring discrimination of binary classification, integers from $0$ to $3$, integers from $0$ to $10$ with the corresponding prompts in Appendix~\ref{app:prompts}.
After QwQ-32B generation, we extract the output score $s_i$ with its probability $\text{Pr}(s_i)$ for computing the final score as discussed in Section~\ref{sec:SFT}.
Table~\ref{tab:discrimination} reports the nDCG@10 averaged on all subsets for each benchmark.

\section{Training Dataset}
\label{app:train_data}

For the hard query dataset (HQ) built by ReasonIR~\cite{ReasonIR}, we randomly sample $10,000$ queries.
In the original dataset, each query has a positive document drawn from the high-quality corpus~\cite{BRIGHT}, and a synthetic hard negative document.
To enrich the negative documents, we use the ReasonIR-8B retriever~\cite{ReasonIR} to retrieve top $1,000$ negative documents from the same corpus, among which the top $10$ are considered as hard negatives.
Also, we randomly sample $4$ documents from positions $11-100$ as medium negatives, and $4$ documents from positions $101-1,000$ as easy negatives.
In this way, we obtain $10,000$ queries and each query is associated with $20$ documents, including one positive document and $19$ negative documents.

For the Promptriever training set~\cite{promptriever}, we randomly sample $5,000$ queries, which contain curated instructions that impose additional requirements for relevance judgment.
In the original dataset, each query has a positive document drawn from the MS MARCO corpus~\cite{MS-MARCO}, and $1$ to $3$ synthetic negative documents.
To enrich the negative documents, we use the ReasonIR-8B retriever~\cite{ReasonIR} to retrieve top $1,000$ negative documents from the same corpus, among which the top $10$ are considered as hard negatives.
For the remaining negative documents, we randomly sample half from positions $11-100$ as medium negatives, and the other half from positions $101-1,000$ as easy negatives.
In this way, we obtain $5,000$ queries and each query is associated with $20$ documents, including one positive document and $19$ negative documents.

For the MS MARCO traing dataset~\cite{MS-MARCO}, we randomly sample $5,000$ queries, each of which only has one positive documents.
Similarly, we use the ReasonIR-8B retriever~\cite{ReasonIR} to retrieve top $1,000$ negative documents from the same corpus, among which the top $10$ are considered as hard negatives.
For the remaining negative documents, we randomly sample half from positions $11-100$ as medium negatives, and the other half from positions $101-1,000$ as easy negatives.
In this way, we obtain $5,000$ queries and each query is associated with $20$ documents, including one positive document and $19$ negative documents.

Finally, we filter out any instances where the generated output exceeds $2,048$ tokens. This procedure results in our final Supervised Fine-Tuning (SFT) dataset $\mathcal{D}$ containing $14,799$ queries, as summarized in Table~\ref{tab:training_data}.
For RL training, we randomly sample $2,048$ queries from the SFT dataset $\mathcal{D}$.

\section{Prompts}
\label{app:prompts}

For baseline methods, we use prompts from their papers and official repositories, with the corresponding token limits for query and document truncation.
For our method and preliminary experiments, queries or documents longer than $2,048$ tokens will be truncated.

\begin{figure}[t]
\begin{prompt}
\small
Given a query and a document, please give a relevance judgement of yes/no.
The goal or relevance definition is: \texttt{\{instruction\}}
\vspace{2mm}

Here is the query:\\
\texttt{\{query\}}
\vspace{2mm}

Here is the document:\\
\texttt{\{doc\}}
\vspace{2mm}

Please directly choose a relevance judgement from [yes, no].
Only output one word, no other words are allowed.
\vspace{2mm}

Your output:
\end{prompt}
\caption{Prompt for Qwen3-32B using binary outputs.}
\label{fig:Qwen3_prompt}
\end{figure}

\begin{figure}[t]
\begin{prompt}
\small

Given a query and a document, please give a relevance judgement of yes/no.
The goal or relevance definition is: \texttt{\{instruction\}}
\vspace{2mm}

Here is the query:\\
\texttt{\{query\}}
\vspace{2mm}

Here is the document:\\
\texttt{\{doc\}}
\vspace{2mm}

After thinking, please directly choose a relevance judgement from [yes, no].
\vspace{2mm}

Desired output format:\\
\texttt{<think>}put your thinking here\texttt{</think>}\texttt{<answer>} Only allows yes/no here\texttt{</answer>}
\vspace{2mm}

Your output:
\end{prompt}
\caption{Prompt for QwQ-32B using binary outputs.}
\label{fig:prompt_binary}
\end{figure}

\begin{table*}[t]
\small
\centering
\begin{tabular}{cp{135mm}}
\toprule
\textbf{Benchmark} (Subset) & \textbf{Instruction} \\
\midrule
BRIGHT (AoPS) & We want to find different but similar math problems to the query. A document is relevant if it uses the same class of functions and shares any overlapping techniques. \\
\midrule
BRIGHT (LeetCode) & I am looking to find different problems that share similar data structures (of any kind) or algorithms (e.g. DFS, DP, sorting, traversals, etc.). I am looking for problems that share one or both of these similarities to the query. Does the passage below share any similarities? e.g. if there was a textbook on leetcode problems, this would be in the same book even though it could be in a different chapter. \\
\midrule
BRIGHT (Pony) & I will use the programming language pony. But to solve the problem above, I need to know things about pony. A passage is relevant if it contains docs that match any part (even basic parts) of the code I will have to write for the above program.\\
\midrule
BRIGHT (TheoremQA-Q) & We want to find a document which uses the same mathematical process as the query. A document is relevant if it uses the same mathematical process as the query.\\
\midrule
BRIGHT (TheoremQA-T) & We want to find a document which uses the same mathematical process as the query. A document is relevant if it uses the same mathematical process as the query. \\
\midrule
BRIGHT (others) & A document is relevant if it contains information that helps answer or address the query. A document is not relevant if it doesn't contain information that helps answer the query, even if it mentions similar topics. \\
\midrule
BEIR / TREC DL & Given a query, retrieval relevant passage. \\
\midrule
FollowIR & Retrieval the relevant passage for the given query. Be careful about the extra requirements about relevance in the query. \\
\bottomrule
\end{tabular}
\caption{Task-specific instruction used in prompts.}
\label{tab:instruction}
\end{table*}

\begin{figure}[t]
\begin{prompt}
\small
Given a query and a document, please give a relevance score of 0 to 3.
The goal or relevance definition is: \texttt{\{instruction\}}
\vspace{2mm}

Here is the query:\\
\texttt{\{query\}}
\vspace{2mm}

Here is the document:\\
\texttt{\{doc\}}
\vspace{2mm}

After thinking, directly choose a relevance score from [0, 1, 2, 3].\\
- 0 represents completely not related.\\
- 3 means perfectly related.
\vspace{2mm}

Desired output format:\\
\texttt{<think>}put your thinking here\texttt{</think><answer>} Only allows an integer here\texttt{</answer>}
\vspace{2mm}

Your output:
\end{prompt}
\caption{Prompt for scoring with integers from 0 to 3.}
\label{fig:prompt_0_3}
\end{figure}

For prompts shown in Figure~\ref{fig:prompt_0_10} and Figures~\ref{fig:Qwen3_prompt}-\ref{fig:prompt_0_3}, we use different instructions listed in Table~\ref{tab:instruction} for different subsets due to the diverse definitions of relevance, many of which are adapted from ReasonIR paper~\cite{ReasonIR}.
Notably, when generating trajectories for training data in Section~\ref{sec:SFT}, the instructions used for ReasonIR hard query (HQ) training set~\cite{ReasonIR} are actually those used in BRIGHT benchmark.
The instruction for MS MARCO training set~\cite{MS-MARCO} is the one for TREC DL benchmark, while the instruction for Promptriever training set~\cite{promptriever} is the one for FollowIR benchmark.

\begin{table}[t]
 \centering
 \setlength{\tabcolsep}{2.6mm}{
 \begin{tabular}{l|ll}
 \toprule
 & \textbf{Configuration} & \textbf{Value} \\
 \midrule
 \multirow{10}{1.8cm}{Shared} & finetuning\_type & lora \\
 & lora\_rank & 32 \\
 & lora\_alpha & 64 \\
 & lora\_target & all \\
 & cutoff\_len & 2048 \\
 & learning\_rate & 1e-4 \\
 & num\_train\_epochs & 1.0 \\
 & lr\_scheduler\_type & cosine \\
 & warmup\_ratio & 0.05 \\
 & bf16 & true \\
 \midrule
 \multirow{4}{1.8cm}{For Qwen3-4B-Base} & template & default \\
 & per\_device\_train\_batch\_size & 8 \\
 & gradient\_accumulation\_steps & 2 \\
 & num\_gpus & 8 \\
 \midrule
 \multirow{4}{1.8cm}{For Qwen3-14B-Base} & template & default \\
 & per\_device\_train\_batch\_size & 1 \\
 & gradient\_accumulation\_steps & 8 \\
 & num\_gpus & 16 \\
 \midrule
 \multirow{4}{1.8cm}{For Qwen3-32B} & template & qwen3 \\
 & per\_device\_train\_batch\_size & 1 \\
 & gradient\_accumulation\_steps & 8 \\
 & num\_gpus & 16 \\
 \bottomrule
 \end{tabular}
 }
 \caption{SFT configurations used in LLaMA-Factory, while those not mentioned are kept as the default values.}
 \label{tab:sft_paras}
\end{table}

\section{SFT Settings}
\label{app:SFT}

We use LLaMA-Factory~\cite{llamafactory} to fine-tune Qwen3-4B-Base, Qwen3-14B-Base, and Qwen3-32B LLMs~\cite{Qwen3} on NVIDIA A100 (80G) GPUs.
We apply LoRA~\cite{lora} on all parameters with rank $32$ and alpha $64$, utilizing the effective batch size of $128$.
Detailed parameters are listed in Table~\ref{tab:sft_paras}.

\begin{table}[t]
 \centering
 \begin{tabular}{l|ll}
 \toprule
 & \textbf{Configuration} & \textbf{Value} \\
 \midrule
 \multirow{11}{20mm}{Shared} & train batch size & 1280 \\
 & max\_prompt\_length & 1024 \\
 & max\_response\_length & 1024 \\
 & learning\_rate & 1e-6 \\
 & mini\_batch\_size & 320 \\
 & clip\_ratio ($\epsilon$) & 0.2 \\
 & use\_kl\_loss & True \\
 & kl\_loss\_coef ($\beta$) & 0.001 \\
 & kl\_loss\_type & low\_var\_kl \\
 & rollout\_n ($G$) & 5 \\
 & gpus\_per\_node & 8 \\
 & total\_epochs & 10 \\
 \midrule
\multirow{2}{20mm}{For 4B LLM} & nnodes & 1 \\
& micro\_batch\_size\_per\_gpu & 20 \\
 \midrule
  \multirow{2}{20mm}{For 14B LLM} & nnodes & 2 \\
 & micro\_batch\_size\_per\_gpu & 10 \\
 \midrule
 \multirow{2}{20.5mm}{For 32B LLM} & nnodes & 2 \\
 & micro\_batch\_size\_per\_gpu & 10 \\
 \bottomrule
 \end{tabular}
 \caption{RL configurations used in verl, and those not mentioned are kept as the default values.}
 \label{tab:rl_config}
\end{table}

\section{RL Settings}
\label{app:RL}

We use the GRPO algorithm~\cite{GRPO} implemented in verl project~\cite{verl} for training on NVIDIA A100 (80G) GPUs.
Detailed detailed configurations are listed in Table~\ref{tab:rl_config}.

\section{Settings on BRIGHT Benchmark}
\label{app:bright_settings}

On BRIGHT benchmark, instead of retrieving by BM25 with original queries, existing studies also use different first-stage retrieval and hybridize with BM25 scores to further improve performance~\cite{Rank1, JudgeRank, Rank-R1-v0.2}.
Thus, we conduct further evaluation with settings described as follows.

\paragraph{First-stage retrieval.}
We include the following settings.
\begin{itemize}[leftmargin=*]
 \item \textit{Retrieve by BM25 using GPT-4 reason-query.}
 The first-stage top $100$ documents are retrieved by BM25 on GPT-4’s CoT reasoning content.
 During reranking phase, all rerankers only access the original queries and candidate documents without using such CoTs.
 \item \textit{Retrieve by ReasonIR-8B using GPT-4 reason-query.}
 Similarly, the documents are retrieved using the GPT-4’s CoT, except for using the state-of-the-art retriever ReasonIR-8B~\cite{ReasonIR}.
 Also, such CoTs are not provided during reranking phase.
\end{itemize}

\paragraph{BM25 Hybrid.}
BM25 hybrid has been widely adopted in recent studies on the BRIGHT benchmark~\cite{JudgeRank, ReasonIR, Rank-R1-v0.2} due to its effectiveness as a low-cost model ensembling strategy. 
These methods combine a reranking score $s_i$ for document $d_i$ with its corresponding BM25 score $s_{\text{BM25}}$. 
\begin{itemize}[leftmargin=*]
\item JudgeRank~\cite{JudgeRank} calculates a final score as $100 \times s_i + s_{\text{BM25}}$.
\item Rank-R1-32B-v0.2~\cite{Rank-R1-v0.2} first applies min-max normalization to reranking and BM25 scores, respectively.
Then, it calculates the final score as $0.1\times$ normalized $s_{\text{BM25}} + 0.9 \times $ normalized $s_i$.
\item We also apply the same strategy as Rank-R1-32B-v0.2 on {\methodname} rerankers, except that we apply Z-score normalization (standardization) to the scores instead of min-max normalization.
After transforming them to have a mean of 0 and a standard deviation of 1, we then calculate the final score as $0.2\times$ normalized $s_{\text{BM25}} + 0.8\times$ normalized $s_i$.
\end{itemize}

\section{Evaluation Results}
\label{app:detailed_results}

Tables~\ref{tab:FollowIR_full}-\ref{tab:BEIR_full} presents the detailed results of each subset on four benchmarks.

\begin{figure}[t]
 \centering
 \includegraphics[width=\linewidth]{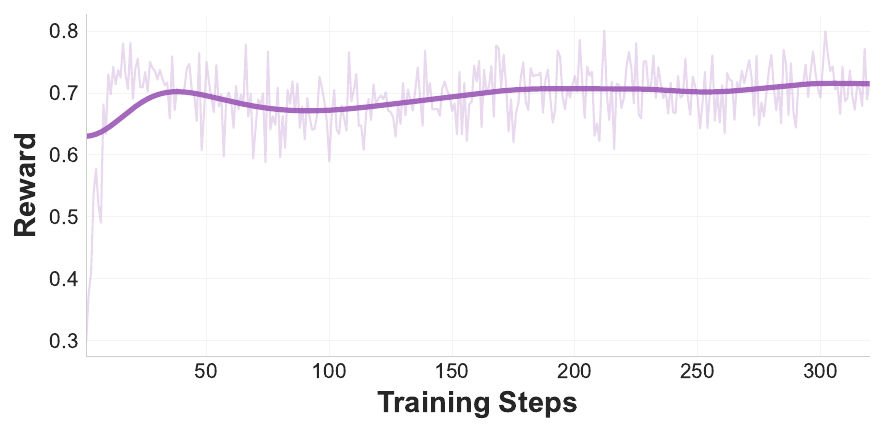}
 \caption{Reward curve during GRPO training for {\methodname}-4B reranker.}
 \label{fig:reward_curves}
\end{figure}

\begin{figure}[t]
 \centering
 \includegraphics[width=\linewidth]{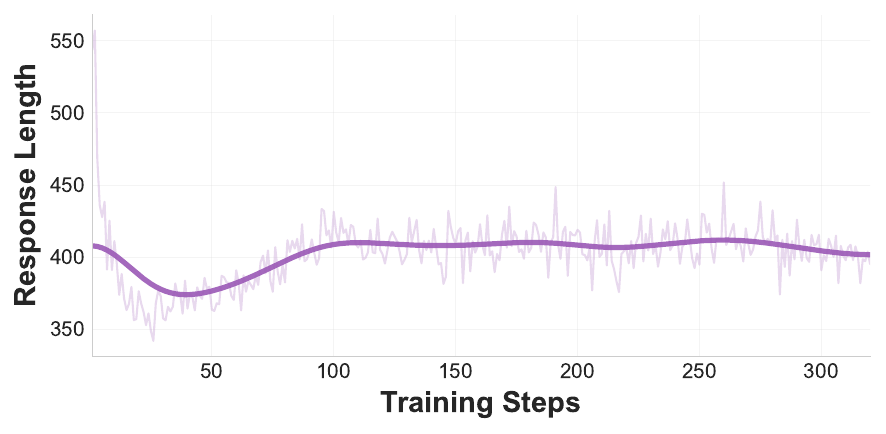}
 \caption{Response length curve during GRPO training for {\methodname}-4B reranker.}
 \label{fig:response_curves}
\end{figure}

We also present the reward and response length curves for the {\methodname}-4B reranker during GRPO training.
As detailed in Appendix~\ref{app:RL}, the training process consists of $10$ epochs, with $32$ steps per epoch. Figure~\ref{fig:reward_curves} shows that the reward initially increases rapidly. After a brief dip from its peak around step $25$, it grows steadily and saturated after step $160$. In Figure~\ref{fig:response_curves}, the response length first drops sharply, then gradually increased, saturating at approximately $410$ tokens after $100$ steps.

\section{Rule-based Rewards}
\label{app:rewards}

In Section~\ref{sec:ablation}, we compare different rule-based reward functions during GRPO training of rerankers.

\paragraph{Pointwise reward $r_\text{SE}$.}
This reward assesses each query-document pair $(q, d_i)$ independently as follows.
\begin{equation*}
r_\text{SE} =
 \begin{cases}
 1 - \frac{(s_i - t_i)^2}{100}, & \text{formatted}, \\
 -1, & \text{otherwise},
 \end{cases}
\end{equation*}
where $s_i$ is the score given by policy model for pair $(q, d_i)$, and $t_i$ is the corresponding reference score from the teacher model, i.e., QwQ-32B~\cite{QwQ} in this paper.
``Formatted'' indicates that the output follows desired format.

This reward motivates the policy model to generate outputs in the correct format and with scores closely matching the reference scores.

\paragraph{Listwise reward $r_\text{nDCG}$.}
Similar to $r_\text{RR}$ reward discussed in Section~\ref{sec:RL}, $r_\text{nDCG}$ is also a listwise reward utilizing all $N \times G$ scores from $G$ rollouts for $N$ documents corresponding to a query. 
For the $j$-th rollout of document $d_i$, let $\phi_i^{(j)}$ denote its rank among the $N \times G$ scores, with ties resolved by assigning the minimum rank associated with the tied ones.

Denote the positive and negative document sets for query $q$ as $\mathcal{D}^P$ and $\mathcal{D}^N$, respectively. 
We first present the Discounted Cumulative Gain (DCG) on $N \times G$ scores:
\begin{equation*}
 \text{DCG} = \sum_{i=1}^{N} \sum_{j=1}^G \frac{2^{rel_i} - 1}{\log_2\left(\phi_i^{(j)}+1\right)},
\end{equation*}
where $rel_i$ is the relevance score of document $d_i$. For simplicity, we assume $rel_i=1$ for positive documents ($d_i \in \mathcal{D}^P$) and $rel_i=0$ for negative ones ($d_i \in \mathcal{D}^N$).
Let $\mathbb{I}(d_i \in \mathcal{D}^P)$ be an indicator function that equals $1$ only when $d_i \in \mathcal{D}^P$, and define $f(\phi_i^{(j)}) = \frac{1}{\log_2\left(\phi_i^{(j)}+1\right)}$.
We have
\begin{equation*}
 \text{DCG} = \sum_{i=1}^{N} \sum_{j=1}^G \mathbb{I}(d_i \in \mathcal{D}^P) f(\phi_i^{(j)}).
\end{equation*}

The Ideal Discounted Cumulative Gain (IDCG) is computed by considering the DCG metric when all positive documents are ranked ahead of all negative ones. Using IDCG, the nDCG metric for $N \times G$ scores is calculated as:
\begin{equation*}
 \text{nDCG} = \frac{\text{DCG}}{\text{IDCG}}.
\end{equation*}

For the $j$-th rollout score of a positive document $d_i \in \mathcal{D}^P$, its contribution to the overall nDCG metric is $\frac{f(\phi_i^{(j)})}{\text{IDCG}}$.
Inspired by this observation, we introduce the $r_\text{nDCG}$ reward as follows. 
Let $\Phi_{\mathcal{D}^P} = \{\phi_i^{(j)} \mid d_i \in \mathcal{D}^P, j\in\{1,2, \cdots, G\}\}$ denote the ranks of all positive documents. 
We denote the maximum and minimum ranks within this set as $\Phi_{\mathcal{D}^P}^{\text{(max)}}$ and $\Phi_{\mathcal{D}^P}^{\text{(min)}}$, respectively. 
The $r_\text{nDCG}$ reward is defined as:
{\small
\begin{equation*}
 r_\text{nDCG} =
 \begin{cases}
 \frac{f(\phi_i^{(j)})}{IDCG}, & \text{formatted, } d_i \in \mathcal{D}^P, \\
 -\frac{f(\Phi_{\mathcal{D}^P}^{\text{(min)}})}{IDCG}, & \text{formatted, } d_i \in \mathcal{D}^N, \phi_i^{(j)} \leq \Phi_{\mathcal{D}^P}^{\text{(max)}}, \\
 1 - \frac{(s_i - t_i)^2}{100}, & \text{formatted, } d_i \in \mathcal{D}^N, \phi_i^{(j)} > \Phi_{\mathcal{D}^P}^{\text{(max)}}, \\
 -1, & \text{otherwise},
 \end{cases}
\end{equation*}
}where ``formatted'' indicates the policy model generates response in desired format and the score $s_i^{(j)}$ can be extracted.
And $t_i$ is the reference score from the reference model $\pi_{\text{ref}}$ (i.e., the SFT-tuned model).

\begin{table}[t]
\centering
\setlength{\tabcolsep}{1.6mm}{
\begin{tabular}{l|c|ccc}
\toprule
\textbf{Method} & \textbf{Average} & Core17 & News21 & Robust04 \\
\midrule
BM25 & 0.0 & 0.0 & 0.0 & 0.0 \\
JudgeRank-8B & 9.9 & 14.6 & 2.0 & 13.1 \\
Rank-R1-7B & 3.6 & 6.1 & 4.0 & 0.7 \\
Rank1-7B & 9.1 & 10.1 & 4.8 & 12.3 \\
Rearank-7B & 2.3 & 4.3 & 0.1 & 2.5 \\
QwQ-32B & 14.1 & 20.9 & 9.4 & 12.0 \\
{\methodname}-4B & 11.0 & 17.5 & 1.3 & 14.2 \\
{\methodname}-14B & 10.3 & 13.6 & 2.0 & 15.4 \\
{\methodname}-32B & 12.1 & 19.0 & 1.7 & 15.5 \\
\bottomrule
\end{tabular}
}
\caption{Detailed $p$-MRR on each subset of FollowIR benchmark.}
\label{tab:FollowIR_full}
\end{table}

\begin{table}[t]
\centering
\setlength{\tabcolsep}{1.8mm}{
\begin{tabular}{l|c|cc}
\toprule
\textbf{Method} & \textbf{Average} & TREC DL19 & TREC DL20 \\
\midrule
BM25 & 49.3 & 50.6 & 48.0 \\
JudgeRank-8B & 62.6 & 65.4 & 59.9 \\
Rank-R1-7B & 70.0 & 72.2 & 67.7 \\
Rank1-7B & 67.1 & 69.0 & 65.1 \\
Rearank-7B & 72.5 & 74.8 & 70.1 \\
QwQ-32B & 66.1 & 67.5 & 64.8 \\
{\methodname}-4B & 68.9 & 71.1 & 66.7 \\
{\methodname}-14B & 67.1 & 68.9 & 65.4 \\
{\methodname}-32B & 68.1 & 70.8 & 65.5 \\
\bottomrule
\end{tabular}
}
\caption{Detailed nDCG@10 on each subset of TREC DL benchmark.}
\label{tab:DL_full}
\end{table}

\begin{table*}[t]
\small
\centering
\setlength{\tabcolsep}{2mm}{
\begin{tabular}{l|c|ccccccc|cc|ccc}
\toprule
\multirow{2}{*}{\textbf{Method}} & \multirow{2}{*}{\textbf{Avg.}} & \multicolumn{7}{c|}{\textbf{StackExchange}} & \multicolumn{2}{c|}{\textbf{Coding}} & \multicolumn{3}{c}{\textbf{Theorem-based}} \\
 &  & Bio. & Earth. & Econ. & Psy. & Rob. & Stack. & Sus. & Leet. & Pony & AoPS & TheoQ. & TheoT. \\
\midrule
\multicolumn{14}{c}{\textit{Retrieve top 100   documents by BM25, using original query}} \\
\midrule
BM25 & 13.7 & 18.2 & 27.9 & 16.5 & 13.4 & 10.9 & 16.3 & 16.1 & 24.7 & 4.3 & 6.5 & 7.3 & 2.1 \\
Rank-R1-7B & 15.7 & 23.4 & 29.2 & 16.4 & 23.0 & 17.0 & 10.9 & 25.9 & 15.8 & 4.8 & 5.8 & 7.1 & 9.3 \\
Rank1-7B & 18.2 & 31.6 & 34.4 & 18.0 & 23.5 & 16.7 & 18.6 & 22.9 & 20.1 & 9.4 & 4.5 & 9.4 & 9.9 \\
Rearank-7B & 17.4 & 23.2 & 26.7 & 17.2 & 22.7 & 18.2 & 16.7 & 25.3 & 26.8 & 7.2 & 7.5 & 7.7 & 9.7 \\
JudgeRank-8B & 17.0 & 28.7 & 32.2 & 20.9 & 24.6 & 16.5 & 18.3 & 20.6 & 11.7 & 7.1 & 4.7 & 8.4 & 10.0 \\
\textit{+ BM25 hybrid} & 19.0 & 28.3 & 36.5 & 21.9 & 24.1 & 15.3 & 22.7 & 23.5 & 25.1 & 6.8 & 6.7 & 8.3 & 8.5 \\
QwQ-32B & 23.2 & 32.7 & 44.7 & 23.9 & 30.5 & 21.6 & 23.8 & 23.8 & 25.7 & 17.3 & 12.7 & 11.2 & 11.1 \\
\textit{+ BM25 hybrid} & 23.9 & 33.0 & 46.5 & 25.3 & 28.2 & 21.1 & 25.6 & 25.3 & 28.7 & 17.2 & 13.0 & 11.8 & 10.7 \\
{\methodname}-4B & 22.7 & 30.4 & 42.5 & 21.5 & 27.7 & 22.4 & 22.9 & 24.0 & 31.6 & 14.6 & 11.0 & 12.1 & 11.4 \\
\textit{+ BM25 hybrid} & 23.9 & 32.7 & 45.4 & 23.1 & 29.2 & 21.8 & 24.7 & 25.6 & 33.4 & 15.6 & 12.2 & 12.4 & 10.5 \\
{\methodname}-14B & 23.1 & 31.2 & 43.6 & 25.8 & 27.8 & 23.1 & 23.9 & 24.6 & 29.8 & 16.8 & 8.6 & 10.5 & 11.9 \\
\textit{+ BM25 hybrid} & 24.6 & 32.7 & 45.8 & 27.2 & 29.4 & 24.1 & 25.6 & 26.5 & 32.7 & 17.5 & 10.5 & 12.1 & 11.9 \\
{\methodname}-32B & 24.4 & 33.5 & 44.5 & 23.9 & 29.4 & 23.8 & 27.1 & 26.4 & 32.5 & 15.8 & 12.5 & 10.9 & 12.2 \\
\textit{+ BM25 hybrid} & 25.4 & 35.1 & 46.2 & 25.5 & 29.4 & 24.2 & 27.5 & 27.6 & 34.9 & 16.7 & 13.2 & 11.5 & 12.5 \\
\midrule
\multicolumn{14}{c}{\textit{Retrieve top 100   documents by BM25, using GPT4 reason-query}} \\
\midrule
BM25 & 27.0 & 53.6 & 54.1 & 24.3 & 38.7 & 18.9 & 27.7 & 26.3 & 19.3 & 17.6 & 3.9 & 19.2 & 20.8 \\
Rank-R1-7B & 23.9 & 38.2 & 29.4 & 23.4 & 33.0 & 24.9 & 14.9 & 33.2 & 18.2 & 16.1 & 3.8 & 16.6 & 34.8 \\
Rank1-7B & 25.5 & 45.8 & 37.0 & 22.2 & 31.7 & 20.6 & 23.0 & 34.2 & 15.7 & 19.8 & 1.3 & 19.8 & 34.7 \\
Rearank-7B & 29.1 & 42.0 & 37.5 & 26.4 & 39.1 & 25.0 & 25.1 & 32.6 & 26.2 & 29.2 & 5.9 & 28.0 & 32.2 \\
XRR-Gemini-2.5-Flash* & 40.3 & 63.1 & 55.4 & 38.5 & 52.9 & 37.1 & 38.2 & 44.6 & 21.9 & 35.0 & 15.7 & 34.4 & 46.2 \\
JudgeRank-8B & 24.4 & 41.4 & 34.7 & 26.2 & 36.0 & 24.0 & 27.6 & 26.1 & 10.2 & 14.2 & 3.4 & 20.3 & 28.9 \\
\textit{+ BM25 hybrid} & 31.0 & 55.3 & 53.4 & 31.4 & 41.6 & 26.7 & 32.8 & 33.3 & 19.6 & 19.5 & 3.7 & 23.4 & 30.9 \\
{\methodname}-4B & 32.9 & 48.2 & 46.7 & 30.0 & 43.1 & 28.4 & 31.5 & 38.1 & 28.5 & 23.5 & 10.4 & 26.9 & 39.0 \\
\textit{+ BM25 hybrid} & 36.1 & 58.5 & 55.6 & 32.6 & 47.2 & 30.0 & 34.7 & 40.6 & 28.9 & 25.8 & 11.2 & 28.9 & 39.0 \\
{\methodname}-14B & 33.5 & 51.4 & 48.6 & 30.8 & 41.3 & 26.7 & 35.6 & 39.1 & 27.3 & 26.4 & 10.9 & 25.7 & 37.9 \\
\textit{+ BM25 hybrid} & 36.7 & 59.9 & 57.3 & 34.8 & 46.7 & 29.5 & 36.9 & 41.2 & 29.4 & 28.7 & 10.5 & 28.0 & 38.1 \\
{\methodname}-32B & 34.6 & 55.5 & 49.1 & 30.4 & 44.7 & 27.9 & 35.6 & 40.6 & 29.2 & 24.2 & 10.4 & 27.6 & 40.0 \\
\textit{+ BM25 hybrid} & 37.4 & 62.9 & 57.5 & 33.2 & 48.4 & 30.5 & 36.5 & 42.3 & 32.7 & 25.4 & 10.8 & 28.7 & 40.4 \\
\midrule
\multicolumn{14}{c}{\textit{Retrieve top 100   documents by ReasonIR-8B, using GPT4 reason-query}} \\
\midrule
ReasonIR-8B & 30.5 & 43.5 & 43.0 & 32.8 & 38.9 & 21.1 & 30.6 & 27.3 & 31.6 & 19.6 & 7.3 & 34.1 & 36.7 \\
Rank-R1-7B & 24.1 & 39.3 & 28.1 & 23.9 & 30.0 & 17.3 & 18.1 & 33.2 & 18.6 & 15.0 & 4.2 & 25.4 & 35.7 \\
Rank1-7B & 24.3 & 44.1 & 33.5 & 21.8 & 30.0 & 15.0 & 22.1 & 28.5 & 11.8 & 21.7 & 1.2 & 26.2 & 36.2 \\
Rearank-7B & 27.5 & 35.3 & 29.8 & 25.5 & 35.7 & 19.1 & 20.1 & 32.9 & 29.9 & 20.2 & 6.2 & 36.7 & 38.3 \\
JudgeRank-8B & 20.2 & 37.1 & 27.2 & 19.2 & 28.6 & 11.6 & 19.9 & 22.5 & 10.2 & 10.2 & 3.6 & 22.9 & 29.4 \\
\textit{+ BM25 hybrid} & 22.7 & 40.4 & 28.9 & 22.3 & 35.5 & 14.2 & 23.0 & 25.7 & 11.8 & 10.6 & 3.6 & 25.2 & 31.1 \\
Rank-R1-v0.2-32B* & 37.7 & 60.1 & 56.3 & 36.6 & 52.1 & 30.2 & 37.6 & 45.9 & 25.5 & 14.6 & 10.1 & 38.6 & 44.3 \\
\textit{+BM25 hybrid*} & 40.0 & 64.4 & 60.1 & 38.3 & 52.2 & 30.7 & 40.6 & 46.7 & 33.3 & 17.4 & 10.1 & 38.6 & 47.7 \\
{\methodname}-4B & 30.5 & 42.1 & 42.5 & 26.3 & 36.4 & 20.8 & 27.3 & 33.2 & 31.7 & 21.8 & 10.9 & 32.8 & 40.6 \\
\textit{+ BM25 hybrid} & 38.7 & 58.7 & 56.6 & 33.8 & 48.7 & 29.1 & 38.2 & 40.8 & 32.7 & 32.0 & 9.8 & 35.2 & 48.4 \\
{\methodname}-14B & 31.8 & 46.6 & 42.5 & 25.2 & 37.3 & 19.6 & 30.2 & 34.6 & 31.9 & 25.6 & 10.5 & 32.4 & 45.0 \\
\textit{+ BM25 hybrid} & 39.3 & 60.1 & 55.8 & 34.2 & 49.5 & 28.4 & 40.1 & 41.2 & 33.3 & 34.0 & 11.0 & 35.7 & 48.5 \\
{\methodname}-32B & 32.8 & 49.3 & 43.4 & 28.4 & 36.8 & 20.8 & 32.8 & 34.6 & 36.0 & 22.3 & 11.3 & 34.4 & 43.5 \\
\textit{+ BM25 hybrid} & 40.2 & 61.5 & 56.6 & 36.5 & 49.4 & 28.9 & 41.8 & 42.7 & 36.0 & 31.6 & 11.4 & 37.0 & 49.1 \\
\bottomrule
\end{tabular}
}
\caption{Detailed nDCG@10 on each subset of BRIGHT benchmark. *Taken from BRIGHT online website.}
\label{tab:BRIGHT_full}
\end{table*}

\begin{table*}[t]
\small
\centering
\setlength{\tabcolsep}{1.18mm}{
\begin{tabular}{l|r|rrrrrrrrrrrrrrr}
\toprule
\textbf{Method} & \textbf{Avg.} & ArguA. & CliF. & CQA. & DBP. & Fever & FiQA & HotP. & MSM. & NFC. & NQ & Quora & SciD. & SciF. & TREC-C. & TouC. \\
\midrule
BM25 & 40.8 & 30.0 & 16.5 & 30.2 & 31.8 & 65.1 & 23.6 & 63.3 & 22.8 & 32.2 & 30.6 & 78.9 & 14.9 & 67.9 & 59.5 & 44.2 \\
JudgeRank-8B & 39.1 & 13.4 & 14.8 & 32.0 & 34.5 & 47.4 & 30.2 & 53.3 & 27.2 & 34.1 & 46.5 & 65.8 & 15.2 & 64.5 & 81.6 & 25.6 \\
Rank-R1-7B & 49.0 & 26.0 & 24.8 & 39.8 & 42.7 & 78.6 & 38.9 & 67.8 & 35.1 & 36.2 & 55.8 & 84.2 & 18.8 & 74.9 & 82.2 & 29.0 \\
Rank1-7B & 44.2 & 20.6 & 16.7 & 33.5 & 37.5 & 66.4 & 37.8 & 62.7 & 31.0 & 35.6 & 53.1 & 70.2 & 17.0 & 76.8 & 79.8 & 25.1 \\
Rearank-7B & 49.0 & 32.1 & 21.9 & 41.0 & 45.2 & 73.8 & 36.6 & 72.3 & 35.5 & 32.7 & 54.5 & 79.1 & 20.2 & 74.8 & 79.7 & 36.1 \\
QwQ-32B & 47.5 & 51.7 & 16.2 & 37.9 & 40.7 & 68.0 & 37.1 & 59.0 & 30.5 & 36.3 & 50.9 & 78.0 & 19.7 & 75.7 & 82.4 & 28.9 \\
{\methodname}-4B & 44.8 & 36.7 & 17.8 & 38.6 & 39.1 & 56.8 & 34.3 & 53.7 & 34.4 & 35.1 & 51.8 & 74.0 & 17.1 & 73.1 & 79.6 & 29.9 \\
{\methodname}-14B & 47.1 & 47.6 & 18.5 & 38.5 & 38.5 & 65.2 & 36.2 & 56.9 & 34.7 & 35.0 & 52.1 & 78.7 & 17.4 & 74.2 & 81.8 & 31.3 \\
{\methodname}-32B & 47.7 & 49.1 & 18.7 & 38.4 & 38.8 & 67.1 & 37.0 & 57.6 & 34.8 & 36.7 & 52.4 & 78.6 & 17.6 & 73.9 & 82.4 & 32.1 \\
\bottomrule
\end{tabular}
}
\caption{Detailed nDCG@10 on each subset of BEIR benchmark.}
\label{tab:BEIR_full}
\end{table*}

\end{document}